\documentclass[useAMS,usenatbib]{mnras}
\pdfoutput=1
\usepackage{aas_macros}
\usepackage{epsfig}
\usepackage{amsmath}
\usepackage{amssymb}
\usepackage{caption}
\usepackage{leftidx}
\usepackage{natbib}
\usepackage{float}

\usepackage[T1]{fontenc}
\usepackage{lmodern}
\usepackage[rgb]{xcolor}
\usepackage{listings}
\definecolor{MyGreen}{rgb}{0.0,0.6,0.3}
\definecolor{MyPurple}{rgb}{0.6,0,0.3}
\usepackage{fancyref}
\usepackage{hyperref}
\hypersetup{colorlinks=true,citecolor=MyGreen,linkcolor=MyPurple,urlcolor=blue}

\title[Centrifugal Mass Loss]{Centrifugally Driven Mass Loss and Outbursts of Massive Stars}

\author[Zhao \& Fuller]{
Xihui Zhao,$^{1,2}$\thanks{E-mail: sherz@mail.ustc.edu.cn}
Jim Fuller$^{2}$
\\
$^{1}$Department of Modern Physics, University of Science and Technology of China, Hefei, Anhui 230026, China\\
$^{2}$TAPIR, Mailcode 350-17, California Institute of Technology, Pasadena, CA 91125, USA
}

\date{Accepted XXX. Received YYY; in original form ZZZ}

\pubyear{2020}

\begin{document}
\label{firstpage}
\pagerange{\pageref{firstpage}--\pageref{lastpage}}
\maketitle

% Abstract of the paper
\begin{abstract}
Rotation and mass loss are crucially interlinked properties of massive stars, strongly affecting their evolution and ultimate fate. Massive stars rotating near their breakup limit shed mass centrifugally, creating Be stars with circumstellar disks and possibly driving outbursts. Using the MESA stellar evolution code, we examine the effects of efficient angular momentum transport on the main sequence and post-main sequence rotational evolution of massive stars. In rapid rotators, angular momentum transported from the contracting core to the expanding envelope can spin up the surface layers past the breakup rate, particularly for stars near (or beyond) the end of the main sequence and in low-metallicity environments. We also demonstrate that centrifugal instabilities could arise in rapidly rotating massive stars, potentially triggering the S Doradus outbursts observed in luminous blue variable stars. Prior mass accretion from a binary companion increases both the likelihood and the intensity of centrifugal mass loss. We discuss implications for massive stellar evolution, Be stars, and luminous blue variables.
\end{abstract}

% Select between one and six entries from the list of approved keywords.
% Don't make up new ones.

%%%%%%%%%%%%%%%%%%%%%%%%%%%%%%%%%%%%%%%%%%%%%%%%%%

%%%%%%%%%%%%%%%%% BODY OF PAPER %%%%%%%%%%%%%%%%%%

\section{Introduction}
\label{intro}

Massive stars lose a significant fraction of their mass over the course of their lives. For single star models, most of this mass loss arises from line-driven winds propelled by photons scattering off absorption lines of iron-group elements (e.g., \citealt{castor:75,vink:12,owocki:14}). However, some massive stars also lose mass through outbursts such as eruptions from luminous blue variable (LBV) stars (e.g., \citealt{humphreys:94,smith:17}). While the physics of line-driven mass loss is largely understood, the mechanisms driving the sporadic outbursts of stars like LBVs are not (see \citealt{smithrev:14} and \citealt{owocki:15} for reviews). 

Rotation is a crucial ingredient in the evolution of massive stars, greatly affecting the appearance, evolution, and mass loss rates (see \citealt{maeder:00,maeder:12} for reviews). For instance, Be stars rotate near their breakup velocity and are surrounded by a disk of centrifugally expelled matter (see review in \citealt{rivinius:13}). In stars near the Eddington limit, the breakup velocity is reduced by radiation force, such that that even modest rotation could potentially significantly increase mass loss rates \citep{friend:86,langer:99,maeder:2000, dwarkadas:02,aerts:04,gagnier:19}. Rotational effects are especially important in low-metallicity stars with weak line-driven winds where AM loss through winds is less important \citep{hirschi:06,georgy:13}. However, many existing studies focus on main-sequence (MS) evolution, and do not incorporate realistic angular momentum (AM) transport in stellar models.

A well-known yet underappreciated aspect of massive stars is that their maximum AM content $J_c$ (assuming rigid rotation) \textit{decreases} strongly as they evolve off the MS. The reason is that massive stars have very large cores and very tenuous envelopes compared to lower mass stars. As massive stars evolve off the MS, the contraction of the core can outweigh the expansion of the envelope, such that the radius of gyration (defined as $\kappa = I/MR^2$, where $I$ is the moment of inertia) can decrease by multiple orders of magnitude (
ure \ref{rotation}). The maximum AM content $J_c$ also decreases, such that a star rotating under breakup on the MS may find itself rotating faster than breakup as it evolves off the MS.

This concept of centrifugally driven mass loss has been explored \citep{langer:98,maeder:01,ekstrom:08} but has not received a great deal of attention, perhaps because its operation depends on the AM transport between the core and the envelope. Without AM transport, the envelope will simply expand and spin down, but it can be forced to rotate faster than breakup in the presence of efficient AM transport. Fortunately, asteroseismology of low-mass stars has clearly demonstrated that an efficient AM transport process is at work (e.g., \citealt{mosser:12,deheuvels:14,cantiello:14,eggenberger:17,vanreeth:18,fuller:19}), motivating new investigations into the rotational evolution of massive stars. 

In this paper, we explore the consequences of centrifugally driven mass loss in stars evolving off the MS, employing AM transport that roughly matches observations of low-mass stars (\citealt{fuller:19}, though see also \citealt{eggenberger:19}). In many cases, ordinary stellar winds remove too much AM during the MS for centrifugally driven mass loss to occur. However, centrifugal mass loss can be very important for moderate-mass and low-metallicity stars with lower mass loss rates. It can also be crucial for stars that have been spun up via accreted mass and AM from a companion star. Furthermore, we begin a preliminary investigation of centrifugally driven instabilities that may drive mass loss through outbursts rather than gradual equatorial or wind mass loss. We discuss implications for stellar evolution, Be stars, and LBVs.

\section{Rotationally Driven Mass Loss}
\label{Rot}

\subsection{Effects of Stellar Evolution}
\label{evolution}

%During the evolution of post-main sequence stars, the cores of stars contract and spin up as a consequence of angular momentum (AM) conservation, then increasing spin rates strengthen the centrifugal acceleration $g_{\rm{rot}}$. At some point, when the break-up limit to the angular velocity $\Omega_{\rm crit}$ is reached by the rotation rate $\Omega$ of the star, the growing centrifugal acceleration $g_{\rm{rot}}$ balances the gravitational acceleration $g_{\rm{grav}}$ and the star then begins to eject mass carrying away angular momentum at its surface.

As stars evolve, their cores typically contract and decrease their moment of inertia, while their envelopes expand and increase their moment of inertia. In the presence of efficient angular momentum (AM) coupling, the surface rotation $\Omega_{\rm surf}$ is then determined by the evolution of the star's total moment of inertia $I$ via $\Omega_{\rm surf} = J/I$, where $J$ is the star's AM. Simultaneously, the star's centrifugal breakup rotation rate scales as $\Omega_{\rm crit} \sim \sqrt{G M/R^3}$ and decreases rapidly as the star's radius $R$ increases. Hence, if the star's mass and AM are conserved, its surface rotation rate can eventually exceed the rotational breakup rate, and the star will be forced to centrifugally expel mass. In later sections, we will account for complications arising from non-rigid rotation, radiation pressure, and mass/AM loss via stellar winds. 

%In Figure \ref{rotation}, we illustrate the rotation structures of three evolving stars to show the trend that stars will spin up and reach the break-up rotation rate. In the $\kappa$ panel of Figure \ref{rotation}, we see $\kappa=I/(MR^2)$ is decreasing sharply, which suggests the radius of the star is expanding much faster than the moment of inertia, and this fact can also be read from the top two panels of Figure \ref{rotation}. As a consequence, the spin rate $\Omega$, inversely proportional to the moment of inertia due to AM conservation, is decreasing much slower than the break-up limit: $\Omega_{\rm crit} \sim \sqrt{GM/R^3}$, thus the star will eventually reach the break-up limit and loose mass. 

\begin{figure}
\includegraphics[scale=0.53]{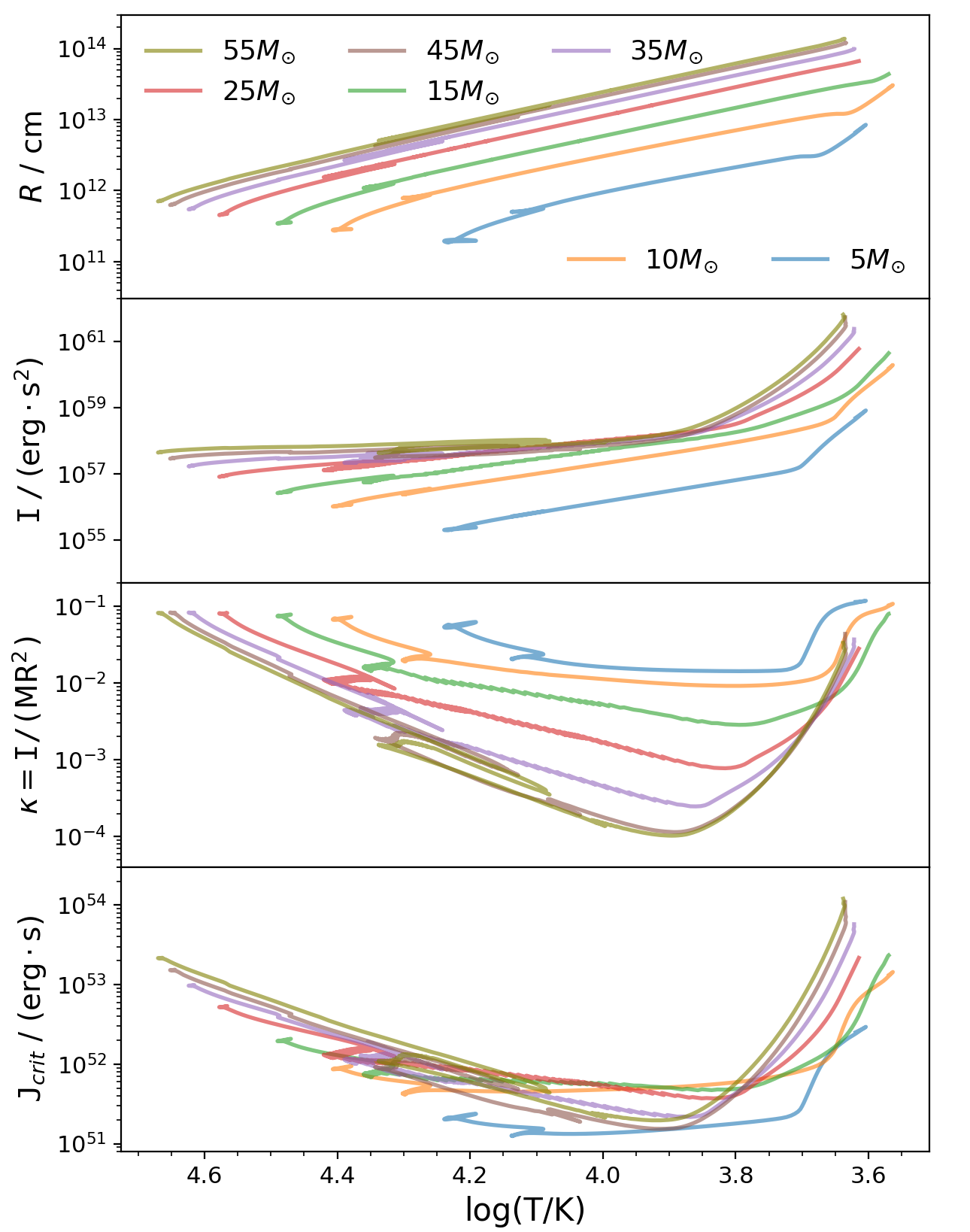}
\caption{\label{rotation}
%Evolution of stellar structure variables related to the star's rotation rate, from the main sequence to core helium burning. All physical quantities are in cgs units, as a function of the logarithm of stellar temperature. {\bf{Panel 1 (top):}} The stellar radius, $R$. {\bf{Panel 2:}} The moment of inertia, $I$. {\bf{Panel 3:}} The dimensionless moment of inertia, $\kappa=I/(MR^2)$. {\bf{Panel 4 (bottom):}} Critical angular momentum $J_{\rm crit}=I\Omega_{\rm crit}$.
Evolution of stellar structure variables related to the star's rotation rate, as a function of surface temperatures, as stars evolve from the main sequence to core helium burning. {\bf{Panel 1 (top):}} The stellar radius, $R$. {\bf{Panel 2:}} The moment of inertia, $I$. {\bf{Panel 3:}} The radius of gyration, $\kappa=I/(MR^2)$. {\bf{Panel 4 (bottom):}} Maximum angular momentum content of a rigidly rotating star, $J_{\rm crit}=I\Omega_{\rm crit}$.
}
\end{figure}

Figure \ref{rotation} illustrates the evolution of stellar properties that determine how their rotation rates change, for several massive star models. We see that stellar radii nearly always increase, while their temperatures decrease, as they evolve through the main sequence, and then across the Hertzsprung--Russell (HR) diagram as they expand into helium-burning red supergiants. The only exception is the small hook at core hydrogen exhaustion, typically at $\log(T_{\rm eff}) \approx 4.3$ for these models. However, we see that the star's moment of inertia $I$ does not always increase strongly, because the expansion of the surface layers is partially compensated by the contraction of the core. This is especially evident for the most massive stars, where the core has a much larger extent on the MS (in both mas and radius), and therefore has a larger influence on the rotational evolution. To demonstrate this, the third panel plots the radius of gyration $\kappa=I/(MR^2)$, which decreases sharply as stars evolve off the MS and become increasingly centrally concentrated.

%As a consequence, the spin rate $\Omega$, inversely proportional to the moment of inertia due to AM conservation, is decreasing much slower than the break-up limit: $\Omega_{\rm crit} \sim \sqrt{GM/R^3}$, thus the star will eventually reach the break-up limit and loose mass.We can also tell such a trend from the bottom panel in Figure \ref{rotation} that demonstrates the maximum angular momentum $J_{\rm crit}=I\Omega_{\rm crit}$ allowed for a rigidly rotating star. $J_{\rm crit}$ is decreasing as the star evolves off the main sequence, so if the total angular momentum $J$ of the star is conserved then at some point, the relation $J>J_{\rm crit}$ will naturally be satisfied and trigger mass loss. Note that $J_{\rm crit}$ drops faster for massive stars, suggesting the fact that this centrifugally driven mass loss effect could be intenser in a more massive star.

Consequently, the star's spin rate $\Omega$, which is inversely proportional to $I$ for a rigidly rotating body, can decrease much slower than the break-up limit, such that stars reach break-up and lose mass. We can determine when this occurs by computing the star's maximum AM content $J_{\rm crit}=I\Omega_{\rm crit}$, shown in the bottom panel of Figure \ref{rotation}. The value of $J_{\rm crit}$ typically decreases by 1-2 orders of magnitude as massive stars cross the HR diagram, such that they can easily reach their breakup limit, $J > J_{\rm crit}$, as they evolve off the MS, simply due to their structural evolution. Because of the large change in $J_{\rm crit}$, this can naturally occur for stars rotating far below their breakup limits on the MS. Again, note that $J_{\rm crit}$ drops faster for more massive stars, suggesting that centrifugally driven mass loss could be more common and more intense for more massive stars. 

%Now we argue for the reliability of our assumption that stars rotate nearly rigidly. While the expanding envelope where mass loss happens actually spins down near the surface of the star because of the AM conservation, surface rotation rate does not slow down very much due to the effective AM transport inside the star, which reduces the internal shear and brings the star closer to rigid rotation. This AM transport couples the envelope to the contracting core and makes rigid rotation a good approximation.

The estimates above assume the star can maintain rigid rotation, which is likely not a good assumption for post-MS stars. In our computations below, in order to compute stars' surface rotation rates, we use asteroseismically calibrated AM transport prescriptions that allow for differential rotation. These calculations thus allow for more realistic predictions of centrifugally driven mass loss rates.

\subsection{Centrifugal Mass Loss Rate}
\label{massloss}

%At the surface of a rotating star, mass loss happens when the centrifugal acceleration $g_{\rm{rot}}$ and gravitational acceleration $g_{\rm grav}$ are balanced, which implies the surface rotation rate $\Omega$ reaches the break-up limit $\Omega_{\rm crit}$. Due to the centrifugal accelerations, the star then ejects mass carrying angular momentum to keep the surface rotation rate at the break-up limit. Here we calculated the mass loss rate resulting from such a centrifugal criterion.

At the surface of a rotating star, mass loss happens when the centrifugal acceleration $g_{\rm{rot}}$ and effective gravitational acceleration $g_{\rm eff}$ are balanced, which implies the surface rotation rate $\Omega_{\rm surf}$ reaches the break-up limit $\Omega_{\rm crit}$. Due to the centrifugal acceleration, the star then ejects mass that carries away AM to keep the surface rotation rate at the break-up limit. To account for this effect, prior works (e.g., \citealt{langer:98,maeder:00}) typically increase the wind mass loss rate when $\Omega_{\rm surf}$ approaches $\Omega_{\rm crit}$. However, the magnitude of this effect is not clear because it depends on the latitudinal surface temperature and wind driving, and some work \citep{owocki:98,muller:14} suggests the wind mass loss rate may actually decrease. We instead choose to estimate the centrifugal mass loss rate purely from the rate at which AM is transported to the surface layers, causing them to be ejected. We note that similar procedures have been performed in \cite{granada:13,georgy:13}.

%Under the assumption that a star rotates rigidly with $J=I\Omega_{\rm crit}$ applied, we derived the relation between mass loss rate and the changing rate of angular momentum of the star, accounting for the fact that mass loss changes the moment of inertia and break-up velocity:
%\begin{align}\label{jdot/j}
%\frac{\dot{J}}{J}=\frac{\dot{\kappa}}{\kappa}+\frac{3}{2}\frac{\dot{M}}{M}+\frac{1}{2}\frac{\dot{R}}{R}
%\end{align}
%where $\kappa=\frac{I}{MR^2}$. Note this relation holds as long as $\Omega_{\rm crit} \propto \sqrt{\frac{M}{R^3}}$. On the other hand, mass loss carries away angular momentum at a rate as: $\dot{J}=\alpha\dot{M}R^2\Omega_{\rm crit}$. Setting them equal, we have the expression for mass loss rate:
%\begin{align}\label{mdot/m}
%\frac{\dot{M}}{M}={\left(\frac{\alpha}{\kappa}-\frac{3}{2}\right)}^{-1}\left(\frac{\dot{\kappa}}{\kappa}+\frac{1}{2}\frac{\dot{R}}{R}\right)
%\end{align}
%This is the mass loss rate corresponding to the centrifugal effect that plays a part when the angular velocity $\Omega$ reaches the break-up limit $\Omega_{\rm crit}$ at the surface of a star.

Assuming that the star rotates rigidly with $J=J_{\rm crit} = I\Omega_{\rm crit} \propto \kappa M^{3/2} R^{1/2}$, the change in AM per unit time is
\begin{align}
\label{jdot/j}
\frac{\dot{J}}{J}=\frac{\dot{\kappa}}{\kappa}+\frac{3}{2}\frac{\dot{M}}{M}+\frac{1}{2}\frac{\dot{R}}{R} \, .
\end{align}
Note this relation holds as long as $\Omega_{\rm crit} \propto \sqrt{M/R^3}$.
On the other hand, mass loss carries away AM at a rate
\begin{equation}
\label{jdot}
\dot{J}=\alpha\dot{M}R^2\Omega_{\rm crit} \, 
\end{equation}
where $\alpha$ accounts for the specific AM of the mass that is lost, in units of $R^2 \Omega_{\rm surf}$. For equatorial mass loss, $\alpha=1$, while $\alpha=2/3$ for spherically averaged mass loss. Setting equations \ref{jdot/j} and \ref{jdot} equal, we have the expression for mass loss rate:
\begin{align}
\label{mdot/m}
\frac{\dot{M}}{M}={\left(\frac{\alpha}{\kappa}-\frac{3}{2}\right)}^{-1}\left(\frac{\dot{\kappa}}{\kappa}+\frac{1}{2}\frac{\dot{R}}{R}\right) \, .
\end{align}
In massive stars, $\kappa \ll 1$ (see Figure \ref{rotation}), so the first term in parentheses dominates, and 
\begin{align}
\label{mdot2}
\frac{\dot{M}}{M} \simeq \frac{\kappa}{\alpha} \left(\frac{\dot{\kappa}}{\kappa}+\frac{1}{2}\frac{\dot{R}}{R}\right) \, .
\end{align}
We only employ this mass loss in our models if $\Omega_{\rm surf}$ approaches $\Omega_{\rm crit}$. While this estimate assumes rigid rotation, which is not strictly valid (especially during post-MS evolution), it has the advantage of being independent of the wind mass loss rate. Equation \ref{mdot/m} may not be precise on a given time step, but it allows mass to be lost at a high rate if $\Omega_{\rm surf} > \Omega_{\rm crit}$, and at a small rate otherwise. Hence, our method is reliable for computing the long-term average mass loss rate when the star evolves such that its surface rotation rate tracks the breakup rotation rate.

In principle, it is possible that the effective value of $\alpha$ could be much larger than unity if a star's AM is transported outward through a circumstellar disk with little corresponding mass loss. We cannot discount such a possibility, so our mass loss rates should perhaps be considered upper limits. However, the AM loss still depends on the production of a disk through centrifugal mass loss, so our predictions for when centrifugal mass loss occurs will not be strongly affected.

%However, in the discussion above we neglect radiative effects, while radiative force in fact can be significant. With Eddington factor $\Gamma$ defined as
%\begin{align}\label{eddfactor}
%\Gamma=\frac{\kappa L}{4\pi cGM},
%\end{align}
%radiative effects neutralize gravitational acceleration when Eddington factor goes to 1, i.e.:
%\begin{align}\label{eddlimit}
%g_{\rm rad}+g_{\rm grav}\rightarrow0 \quad \rm{when} \quad \Gamma\rightarrow 1.
%\end{align}
%It is therefore necessary to add a new criterion implementing radiative effects to the mass loss rate especially when the star is close to the Eddington limit.

\subsection{The Eddington Limit}
\label{eddington}

In the discussion above, we neglected radiative effects, but the radiative force in fact can be large in a massive star, whose Eddington factor $\Gamma$ is defined as
\begin{align}
\label{eddfactor}
\Gamma= \frac{g_{\rm rad}}{g_{\rm grav}} = \frac{\kappa_R L}{4\pi cGM} \, ,
\end{align}
where $L$ is the star's luminosity and $\kappa_R$ is the Rosseland mean opacity at the surface. Radiative forces balance the gravitational force when the Eddington factor approaches unity, i.e.:
\begin{align}
\label{eddlimit}
g_{\rm grav} - g_{\rm rad} \rightarrow0 \quad \rm{when} \quad \Gamma\rightarrow 1.
\end{align}
It is therefore necessary to add a new criterion implementing radiative effects to the mass loss rate, especially when the star is close to the Eddington limit. We shall find below that the centrifugal mass loss timescale is much larger than the stellar expansion timescale, so that the change in $\Gamma$ with mass loss can be neglected when deriving equation \ref{jdot/j} and in equation \ref{mdot/m}. 

%Considering radiative effects, mass loss generally happens when the total acceleration equals to 0 in a rotating star:
%\begin{align}\label{gtot=0}
%{g_{\rm{tot}}}={g_{\rm{grav}}}+{g_{\rm{rot}}}+{g_{\rm{rad}}}={g_{\rm{eff}}}+{g_{\rm{rad}}}=0
%\end{align}
%By using the von Zeipel theorem \citep{vonZeipel:1924} and its generalization \citep{zahn:92,maeder:1999}, \cite{maeder:2000} expressed the total gravity at colatitude $\theta$ as
%\begin{align}\label{gtot}
%{g_{\rm{tot}}=g_{\rm{eff}}}[1-\Gamma_{\Omega}(\theta)]
%\end{align}
%where ${g_{\rm{eff}}}={g_{\rm{rot}}}+{g_{\rm{grav}}}$, and $\Gamma_{\Omega}(\theta)$ is the local Eddington ratio defined as
%\begin{align}\label{localEdd}
%\Gamma_{\Omega}(\theta)=\frac{\kappa(\theta)L }{4\pi c %GM(1-\frac{\Omega^2}{2\pi 
%G\rho_{m}})}
%\end{align}
%Here, $\kappa(\theta)$ is opacity as a function of colatitude $\theta$, $\rho_m$ is the internal average density, $\Omega$ and $L$ are separately the angular velocity and the luminosity on an isobaric surface \citep{zahn:92}. And we assume the deviation of the von Zeipel theorem due to the baroclinicity of the star is very small \citep{maeder:1999,maeder:2000}.

However, the radiative force is important when defining the value of $\Omega_{\rm crit}$ above which centrifugal mass loss occurs. Considering radiative effects, mass loss generally happens when the total acceleration equals zero in a rotating star:
\begin{align}
\label{gtot=0}
{g_{\rm{tot}}}={g_{\rm{grav}}}-{g_{\rm{rot}}}-{g_{\rm{rad}}}={g_{\rm{eff}}}-{g_{\rm{rad}}}=0
\end{align}
By using the von Zeipel theorem \citep{vonZeipel:1924} and its generalization \citep{zahn:92,maeder:1999}, \cite{maeder:2000} expressed the total gravity at colatitude $\theta$ as
\begin{align}
\label{gtot}
{g_{\rm{tot}}=g_{\rm{eff}}}[1-\Gamma_{\Omega}(\theta)]
\end{align}
where ${g_{\rm{eff}}}={g_{\rm{rot}}}+{g_{\rm{grav}}}$, and $\Gamma_{\Omega}(\theta)$ is the local Eddington ratio defined as
\begin{align}
\label{localEdd}
\Gamma_{\Omega}(\theta)=\frac{\kappa_R(\theta)L }{4\pi c GM\big(1-\frac{\Omega^2}{2\pi G\rho_{m}} \big)} \, .
\end{align}
Here, $\rho_m$ is the average density of the star, while $\Omega$ and $L$ are evaluated at isobaric surfaces (i.e., we assume the baroclinicity of the star is very small \citealt{zahn:92,maeder:1999,maeder:2000}).

Equation \ref{gtot=0} has two roots, one is that given by ${g_{\rm{eff}}}=0$, which corresponds to the classical break-up limit:
\begin{align}\label{vcrit1}
\Omega_{\rm{crit,1}} = \sqrt{\frac{2}{3}\frac{GM}{R^3}}
\end{align}
The other root $\Omega_{\rm crit,2}$ corresponds to the condition $\Gamma_{\Omega}(\theta)=1$, and the lower value of these two roots will dominate since mass loss will start as soon as the lower critical velocity is reached. \cite{maeder:2000} argued that only for an Eddington factor $\Gamma$ larger than 0.639, the second root $v_{\rm crit,2}$ has a lower value than $v_{\rm crit,1}$ and is given by:
\begin{align}\label{vcrit2}
v^2_{\rm crit,2}=\frac{9}{4}v_{\rm crit,1}^2\frac{1-\Gamma_{\rm max}}{V'}\frac{R^2_{e}}{R^2_{p}} \, .
\end{align}
Here, $\Gamma_{\rm max}$ is the largest value of the Eddington factor over the stellar surface, $R_{e}$ and $R_{p}$ are the equatorial and polar radius respectively satisfying $R_{e}=\frac{3}{2}R_{p}$ at break-up, and $V'$ is the ratio of the actual volume of a star to the volume of a sphere of radius $R_{p}$, i.e. $V'=V/\frac{4}{3}\pi R^3_{p}$. \cite{maeder:2000} further pointed out that the ratio $v_{\rm crit,2}/v_{\rm crit,1}=1$ when $\Gamma_{max}=0.639$ and $v_{\rm crit,2}/v_{\rm crit,1}=0$ when $\Gamma_{max}=1$. The trend in $v_{\rm crit,2}/v_{\rm crit,1}$ is approximately a linear decrease above $\Gamma = 0.639$. We therefore use the following approximation for the second root:
\begin{align}\label{omegacrit2}
\Omega_{\rm crit,2}=\left(\frac{1-\Gamma}{0.361}\right)\Omega_{\rm crit,1} \quad {\rm when} \quad \Gamma > 0.639 \, .
\end{align}
Since $\Gamma$ is not strongly by centrifigual mass loss, $\Omega_{\rm crit,1}$ and $\Omega_{\rm crit,2}$ both scale as $\Omega_{\rm crit} \propto \sqrt{M/R^3}$, so the mass loss rate formula (\ref{mdot/m}) still applies for the second critical rotation rate $\Omega_{\rm crit,2}$. 

%To consider the scenario in which both radiative and rotational effects have significant contributions to mass loss process, we first set the critical rotation rate $\Omega_{\rm crit}$ to be (\ref{vcrit1}) for $\Gamma<0.639$ and (\ref{omegacrit2}) for $\Gamma>0.639$, since they are separately the roots of the lower value of the equation ${g_{\rm{tot}}}=0$, corresponding to the value of $\Gamma$. Then as soon as $\Omega_{\rm crit}$ is reached by the surface angular velocity of the star, we implement a mass loss rate $\dot{M}$ decided by formula (\ref{mdot/m}) to loose mass.

%To sum up, we set the critical rotation rate $\Omega_{\rm crit}$ to be equation \ref{vcrit1} for $\Gamma<0.639$, and equation \ref{omegacrit2} for $\Gamma>0.639$. Whenever $\Omega_{\rm surf} > \Omega_{\rm crit}$ in our models, we implement a mass loss rate $\dot{M}$ via equation \ref{mdot/m}.

While there has been substantial debate regarding the effect of rotation on massive star wind mass loss rates (e.g., \citealt{langer:98,glatzel:98,owocki:98,maeder:00,gagnier:19}), we emphasize that our predicted mass loss rates do not depend directly on this relationship, but only on whether the star can approach the breakup limit. If rotationally enhanced mass loss of the form advocated by \cite{langer:98} can prevent near-critical rotation from being realized, centrifugal mass loss will not occur. However, in light of recent two-dimensional results \citep{muller:14,gagnier:19}, this appears unlikely, since rotation only enhances wind mass loss very close to the breakup limit.

\section{Results}
\label{sec3}

\subsection{Models}

%Constructing stellar models with the MESA stellar evolution code \citep{paxton:11, paxton:13, paxton:15, paxton:18, paxton:19}, we predict the evolution of rotating massive stars with large initial mass ranging from 10$M_{\odot}$ to 55$M_{\odot}$ and an initial surface rotational velocity of 200 km/s. We study stars with initial solar metallicity $Z=0.017$ and a rotational mixing via Eddington-Sweet circulation with the factor of rotational component of diffusion coefficient for mixing of material equal to $3.33\times 10^{-2}$. The models use the default centrifugal distortion in MESA with limited maximum centrifugal forces, and the angular momentum transport prescription from \cite{fuller:19}. A mechanism implementing mass loss process in rotating stellar models is included in our code as discussed at the end of Section \ref{eddington}, where both radiative and centrifugal effects are significant prescriptions. Here we briefly repeat it.

Constructing stellar models with the MESA stellar evolution code \citep{paxton:11, paxton:13, paxton:15, paxton:18, paxton:19}, we compute the evolution of rotating massive stars with large initial mass ranging from 5$M_{\odot}$ to 55$M_{\odot}$. Our models have an initial surface rotational velocity of 200 km/s, a fairly typical ZAMS value for O/B type stars in the Milky Way and Magellanic clouds \citep{mokiem:06,dufton:06b,hunter:08,ramirez:13,dufton:13}. We will study stars with solar metallicity $Z=0.017$ and LMC metallicity $z=0.0085$. Rotational mixing is included via Eddington-Sweet circulation with the usual rotational mixing factor of $3.33\times 10^{-2}$ (see Appendix \ref{mesa} for a model inlist). The models use the default centrifugal distortion in MESA with limited maximum centrifugal forces, and the AM transport prescription from \cite{fuller:19}.

\begin{figure}
\begin{center}
\includegraphics[scale=0.53]{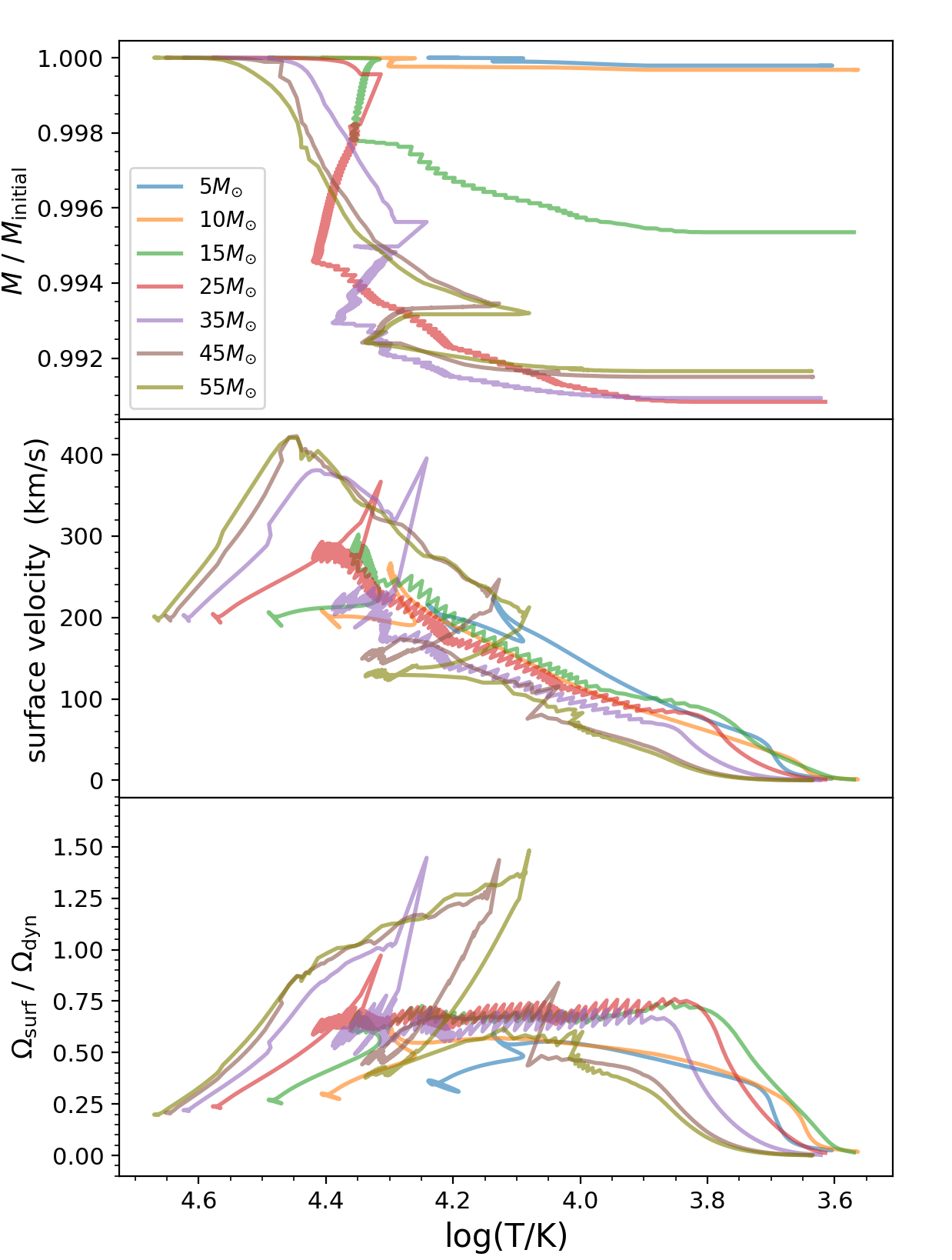}
\end{center}
\caption{\label{mass_nowind}
%{\bf{Top:}} Evolution of the mass of our models including only centrifugal mass loss and no stellar winds. The x-axis is the logarithm of the effective temperature in Kelvin unit of the star. Each model has an initial surface rotational velocity as 200km/s while different colors correspond to different initial masses. Star's lose only a very small fraction of their mass due to centrifugal mass loss, but most of the mass loss occurs during the brief phase when star's cross the Hertzsprung gap (see Figure \ref{HR_nowind}). {\bf{Bottom:}} Plot for surface rotational velocity with the same x-axis as above. Velocities at first speed up quickly and then steadily decrease to zero.
{\bf{Top:}} Evolution of the mass of our models including only centrifugal mass loss and no stellar winds, as a function of their surface temperature. The models begin at the top left and generally evolve towards cooler temperatures. Each model has an initial surface rotational velocity of 200 km/s, with line color indicating initial mass. Stars lose only a very small fraction of their mass due to centrifugal mass loss, but most of the mass loss occurs during the brief phase when stars cross the Hertzsprung gap (see Figure \ref{HR_nowind}). {\bf{Middle:}} Evolution of the surface rotational velocity.  {\bf{Bottom:}} Dimensionless surface spin rates in units of $\sqrt{GM/R^3}$. Note that the dimensionless spin increases to values near unity, at which point centrifugal mass loss enforces $\Omega/\Omega_{\rm dyn} \sim \sqrt{2/3}$ until further expansion of the star reduces the dimensionless spin. 
%For high-mass stars without winds, the surface velocity generally increases until the breakup rate is reached, at which point the star remains rotating near the breakup limit until its velocity further decreases when the star becomes a red supergiant and its moment of inertia increases.
}
\end{figure}

Our models use MESA's MLT++ prescription to handle the structure of super-Eddington near-surface layers. The structure of these layers (typically from the iron opacity peak outward) is dependent on this approximation, as is the radius of the star. Hence, our models do not exhibit the same envelope inflation seen in models without a similar prescription (e.g., \citealt{grafener:12}). We shall discuss uncertainties associated with this issue in Section \ref{discussion}.

%A mechanism implementing mass loss process in rotating stellar models is included in our code as discussed at the end of Section \ref{eddington}, where both radiative and centrifugal effects are significant prescriptions. Here we briefly repeat it. We set the break-up limit $\Omega_{\rm crit}$ to the surface rotation rate of the star to be:
%\begin{equation}
%\Omega_{\rm crit}=\left\{
%\begin{aligned}\label{omegacrit}
%\Omega_{crit,1} & = \sqrt{\frac{2}{3}\frac{GM}{R^3}} \quad , \Gamma<0.639
%\\
%\Omega_{crit.2} & = \left(\frac{1-\Gamma}{0.361}\right)\sqrt{\frac{2}{3}\frac{GM}{R^3}} \quad , \Gamma>0.639
%\end{aligned}
%\right.
%\end{equation}
%and once such limit is reached, the stellar model will then loose mass at a rate as:
%\begin{align}\label{mdot}
%\dot{M}=M\left(\frac{\alpha}{\kappa}-\frac{3}{2}\right)^{-1}\left(\frac{\dot{\kappa}}{\kappa}+\frac{1}{2}\frac{\dot{R}}{R}\right)
%\end{align}
%We run our models until most of the helium in the core has been burned into carbon and oxygen, i.e. the helium mass fraction in the core is below 10\%. 

We include centrifgual mass loss as discussed in Section \ref{eddington}. To summarize, we approximate the surface rotational break-up limit $\Omega_{\rm crit}$ to be:
\begin{equation}
\Omega_{\rm crit}=\left\{
\begin{aligned}\label{omegacrit}
\Omega_{\rm crit,1} & = \sqrt{\frac{2}{3}\frac{GM}{R^3}} \quad , \quad \Gamma<0.639
\\
\Omega_{\rm crit.2} & = \left(\frac{1-\Gamma}{0.361}\right)\sqrt{\frac{2}{3}\frac{GM}{R^3}} \quad , \quad \Gamma>0.639
\end{aligned}
\right.
\end{equation}
For surface rotation rates above $\Omega_{\rm crit}$, we impose a mass loss rate of
\begin{align}\label{mdot}
\dot{M}=M\left(\frac{\alpha}{\kappa}-\frac{3}{2}\right)^{-1}\left(\frac{\dot{\kappa}}{\kappa}+\frac{1}{2}\frac{\dot{R}}{R}\right) \, ,
\end{align}
with $\alpha = 1$. To aid numerical convergence, we smoothly increase mass loss towards equation \ref{mdot} as the critical rotation rate is approached, allowing some centrifugal mass loss at lower rotation rates. We run our models until most of the helium in the core has been burned into carbon and oxygen, i.e. the helium mass fraction in the core is below 10\%.

\begin{figure}
\begin{center}
\includegraphics[scale=0.33]{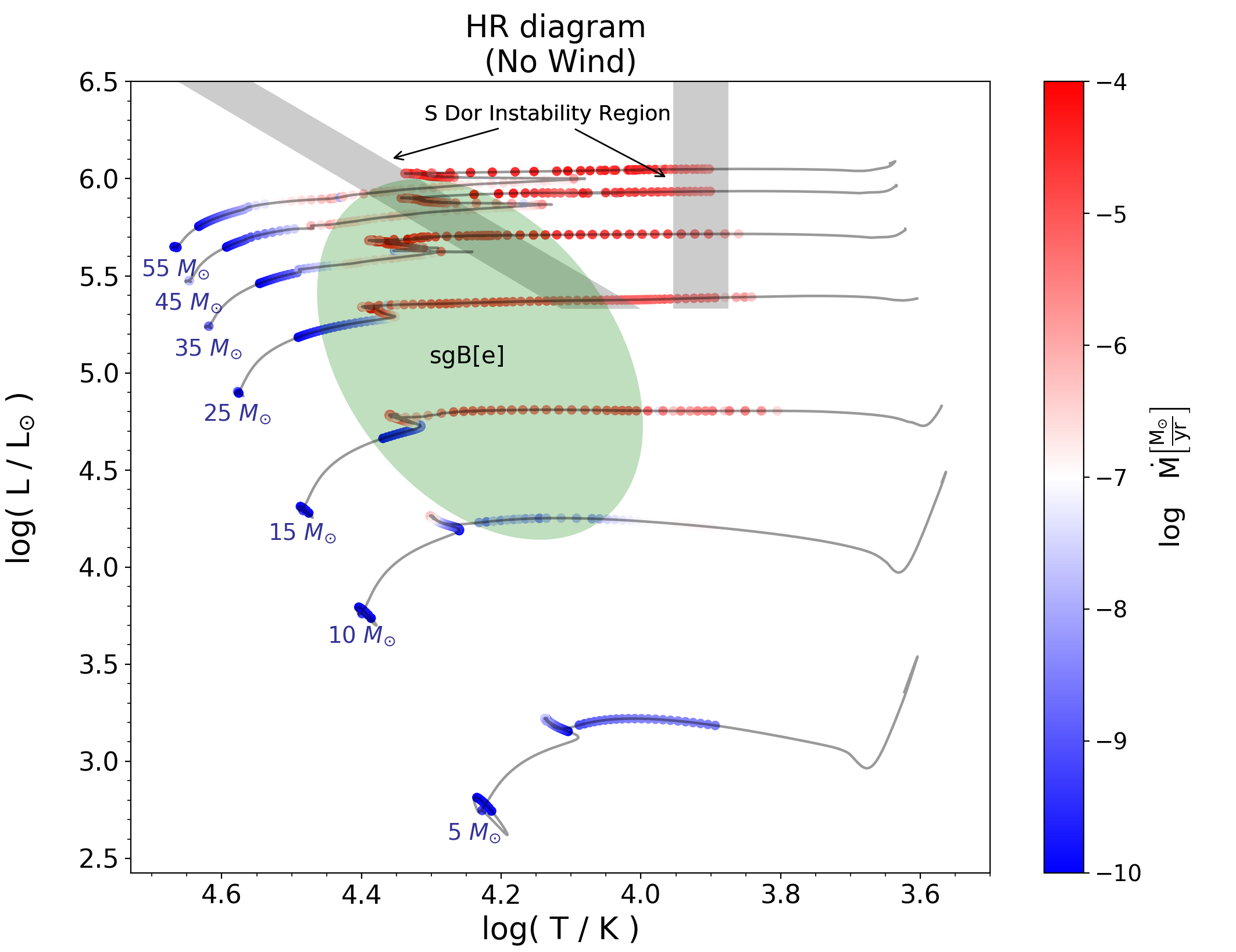}
\end{center}
\caption{\label{HR_nowind}
%Hertzsprung--Russell diagram of stars with initial mass of 10, 15, 25, 35, 45, 55 $M_{\odot}$ and initial surface rotation velocity of 200 km/s, including only centrifugal mass loss and no winds. Points are colored by mass loss rate, and it is evident that large mass loss rates are achieved in the Hertzsprung gap, near the empirical location of the S Doradus instability strip (gray regions) and super-giant B[e] stars (green region). 
Hertzsprung--Russell diagram showing model evolutionary tracks of stars with an initial surface rotation velocity of 200 km/s, including only centrifugal mass loss and no winds. Points are colored by mass loss rate. Large mass loss rates are achieved in the Hertzsprung gap, near the empirical location of S Doradus variability cycles (gray regions) and super-giant B[e] stars (green region) as shown in \citealt{smith:15}.
}
\end{figure}

\subsection{Results without Wind}
\label{nowind}

%In this section we show some results with our own mass loss criterion, while the default Dutch wind prescription in MESA is turned off.

To better understand the effects of stellar evolution and centrifugally driven mass loss, we first construct models without mass loss via radiatively driven winds. These models conserve mass and AM, unless they lose mass centrifugally, and are simpler to understand. While real stars do lose mass and AM through winds, our models in this section are similar to very low-metallicity stars, for which mass loss is irrelevant. They also resemble the post-mass transfer evolution of stars whose AM has been rejuvinated by accretion from a companion.

%Figure \ref{mass_nowind} shows masses (as a fraction of initial mass) and surface rotational velocities of our evolutionary models. Different colors correspond to different initial masses. We see all those stars only lose a small fraction of mass around 1\%, i.e. $\sim 0.1M_{\odot}$, here we deduce an estimation to explain such a small fraction of mass loss.

Figure \ref{mass_nowind} displays the evolution of stellar mass and rotation velocity as a function of effective temperature for our stellar models. During most of the MS, centrifugal mass loss does not occur (or occurs at low levels) because our models are born rotating well below their breakup velocity, and they must evolve substantially before any centrifugal mass loss can occur. During this stage, the stellar mass is nearly constant as stars evolve to cooler temperatures. Stars with larger initial rotational velocities would lose mass earlier along the MS, and could create Oe/Be stars, though the MS mass loss rate is typically modest, as shown in Figure \ref{HR_nowind}.

However, the centrifugal mass loss rate can increase by orders of magnitude near and just after the TAMS, typically at surface temperatures $\log(T) \approx 4.3-3.9$. It is at this stage of evolution that core contraction is fastest, driving centrifugal mass loss at rates up to $10^{-4} \, M_\odot$/yr, with higher mass loss rates for more massive stars. Higher mass models typically lose more total mass than low-mass stars, though none of our models lose more than 1\% of their mass via centrifugal mass loss. Mass loss typically ceases at temperatures of $\log T \lesssim 3.9$, when stars start to develop massive convective envelopes, causing the moment of inertia and $J_{\rm crit}$ to increase sharply, such that stars no longer rotate above their breakup velocities.

%As the beginning of \emph{Section 2}, a star is assumed to rotate rigidly with $J=I\Omega_{\rm crit}$ applied, and we further assume that mass loss happens mostly around the equatorial ring, since equator is in general where the equation ${g_{\rm{tot}}}=0$ is first satisfied, so mass loss carries away angular momentum as:
%\begin{align}\label{deltaJ}
%\Delta J=\Delta M R^2 \Omega_{\rm crit}
%\end{align}
%A star actually loses most of its angular momentum in a small amount of mass loss, i.e. $J\sim \Delta J$, therefore we finally have:
%\begin{align}\label{kappa}
%\kappa \sim \frac{\Delta M}{M}
%\end{align}
%As shown in Figure \ref{rotation}, $\kappa$ is $\sim$ 1\%, which together with (\ref{kappa}) accounts for the small fraction of mass loss. Figure \ref{HR_nowind} is the HR diagram showing the evolutionary paths of our stars. Strong mass ejection generally starts near the Hertzsprung gap. And for massive stars ($M_{initial}>25M_{\odot}$), mass loss stops after they cross the S Dor instability region.

\begin{figure}
\includegraphics[scale=0.61]{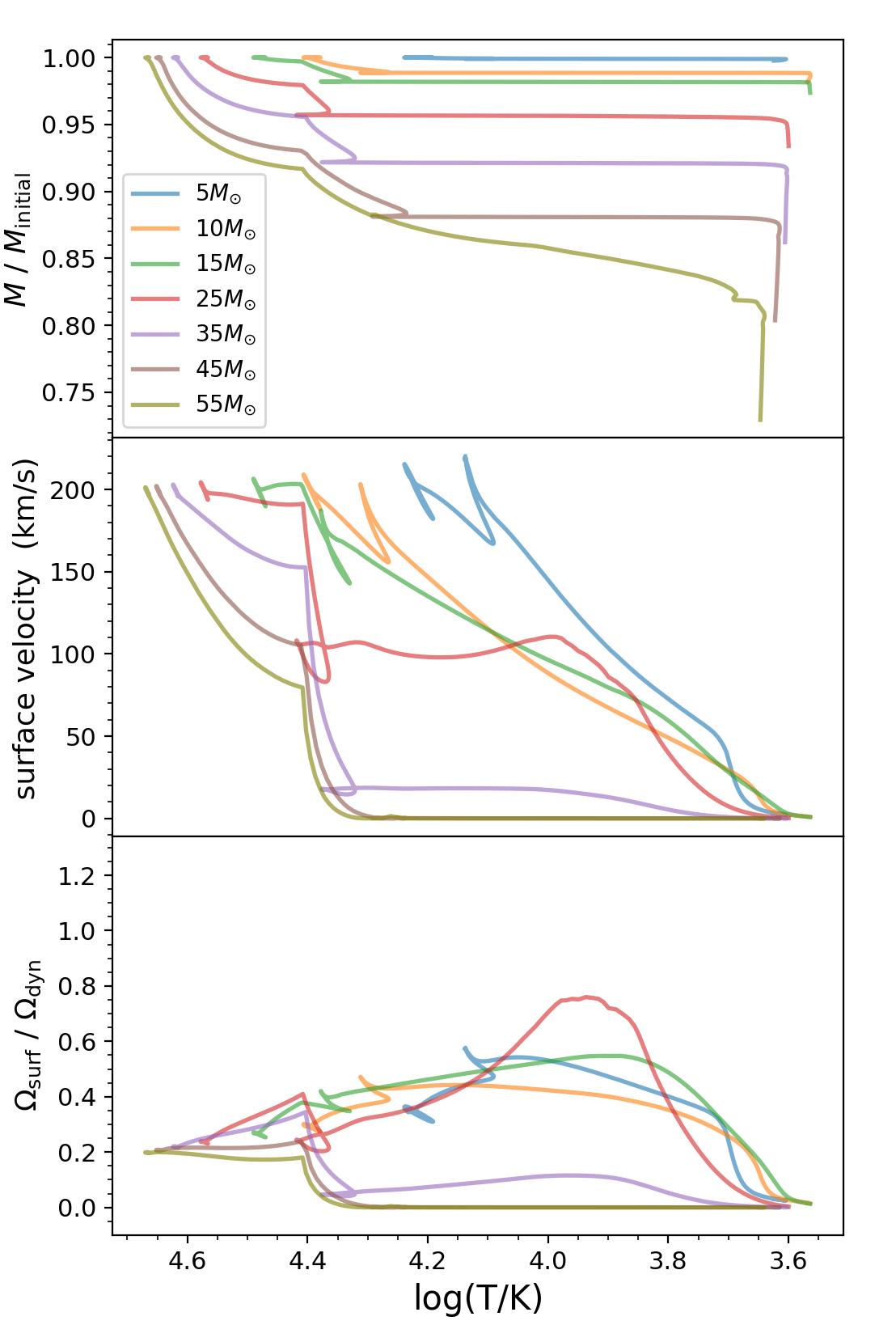}
\caption{\label{mass_wind} Same as Figure \ref{mass_nowind}, but for models at solar metallicity with wind mass loss. These stars generally lose much more mass from winds than from centrifugal mass loss, causing their surface velocities to decrease such that they typically remain rotating slower than their breakup velocities.
}
\end{figure}

The total amount of centrifugal mass loss is easily understood, assuming near rigid rotation.  The ejected mass carries away an AM content of
\begin{align}\label{deltaJ}
\Delta J = \alpha \Delta M R^2 \Omega_{\rm crit}
\end{align}
Hence, in order for a star to lose all of its AM, $J = \kappa M R^2 \Omega_{\rm crit}$, and assuming $\alpha \sim 1$ the amount of mass lost is simply
\begin{align}
\label{kappa}
\frac{\Delta M}{M} \sim \kappa  \, .
\end{align}
As shown in Figure \ref{rotation}, $\kappa \sim 10^{-3}-10^{-2}$ when centrifugal mass loss is occurring, which together with equation \ref{kappa} accounts for the small fraction of mass lost. In a real star with differential rotation, the core rotates faster than the surface and the star has more AM, thus more mass can be lost than predicted by equation \ref{kappa}, but our numerical calculations indicate the total mass loss remains small.

Figure \ref{mass_nowind} also shows the evolution of the surface rotation rate of our windless models. For less massive stars ($M \lesssim 10 \, M_\odot$), the core contraction and envelope expansion roughly balance, such the surface rotation velocity is nearly constant. However, in massive stars, the large contracting core overwhelms the expanding envelope, donating enough AM to increase the surface rotation rate of the star towards breakup, even during the MS. Once centrifugal mass loss begins, the surface velocity tracks the breakup rotation velocity (apart from numerical inaccuracies) until the mass loss ceases at cooler temperatures where the star's maximum allowable AM content $J_c$ increases. At this point, the surface rotation rate approaches zero as the star's moment of inertia becomes very large.

\subsection{Results with wind}

\begin{figure}
\includegraphics[scale=0.34]{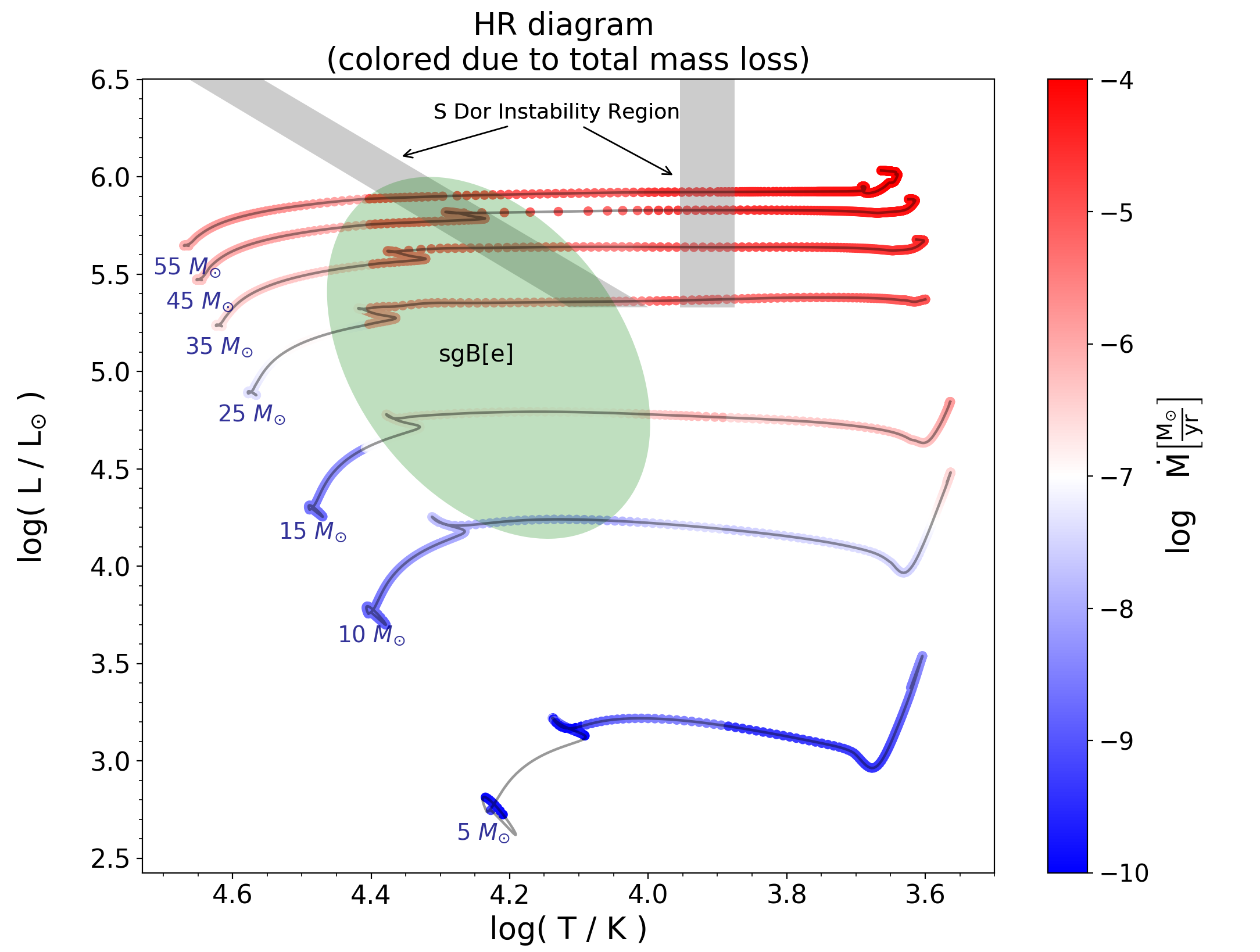}
\hspace{0cm}
\includegraphics[scale=0.34]{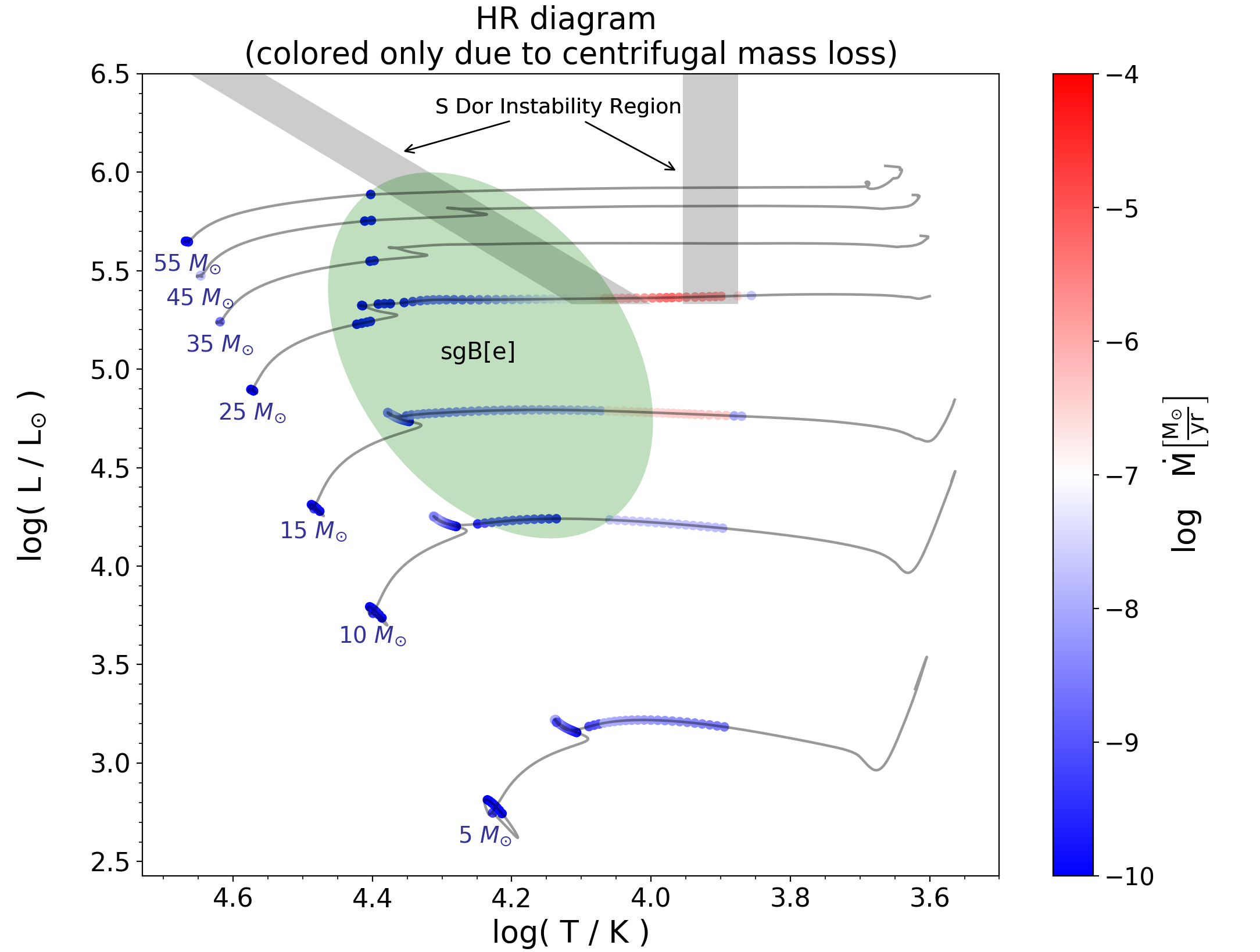}
\caption{\label{HR_wind} Same as Figure \ref{HR_nowind}, for models including both centrifugal mass loss and stellar winds. {\bf{Top:}} Points are colored by total mass loss rate. {\bf{Bottom:}} Points are colored only by centrifugal mass loss, and it is evident that centrifugal effects are insignificant, except for some models crossing the Hertzsprung gap.
}
\end{figure}

%We further examined mass and angular momentum loss of models with both our centrifugal mass loss and mass loss via stellar winds, using the "Dutch wind" prescription in MESA, with a scaling factor of 0.5. We use this value because a reduction factor of 2-3 from prior estimates, due to wind clumping, is necessary to reproduce obervational data (see e.g., \citealt{vink:17,smith:14}).

Real stars lose mass and AM through line-driven winds, complicating the evolution from the simple scenario above. We examine the coupled effects of centrifugal mass loss and wind mass loss using the 'Dutch wind' prescription in our MESA models, with a scaling factor of 0.5. We use this value because a reduction factor of at least 2-3 from prior estimates, due to wind clumping, is necessary to reproduce observational data (see e.g., \citealt{smithrev:14,vink:17}).

%Figure \ref{mass_wind} illustrates the evolution of mass and surface rotational velocity of our models with Dutch wind. Compared to Figure \ref{mass_nowind}, it is obvious that models with wind lose much more mass, and their surface rotational velocities decrease more quickly. Such differences are especially prominent for stars with larger initial mass, indicating that for these models, the centrifugal mass loss is almost irrelevant since the loss in mass and angular momentum caused by wind is much greater.

Figure \ref{mass_wind} illustrates the evolution of mass and surface rotational velocity of our models with wind mass loss. Compared to Figure \ref{mass_nowind}, it is obvious that models with wind lose much more mass, and their surface rotational velocities decrease more quickly. Such differences are especially prominent for stars with larger initial mass because the wind carries away enough AM to prevent the star from ever rotating near breakup. We note the sudden decrease in rotational velocity at $\log(T/{\rm K}) \lesssim 4.4$, due to the bi-stability jump at which the wind mass loss rate increases sharply. Stars with $M \lesssim 10 \, M_\odot$ are not strongly affected due to their low wind mass loss rates. We note that these models never rotate faster than breakup, yet they still lose mass according to our numerical prescription (Appendix \ref{numerics}), which ramps up the mass loss smoothly as the star approaches breakup. The actual threshold of $\Omega_{\rm surf}/\Omega_{\rm dyn}$ for centrifugal mass loss in Be stars is still debated, but is likely significantly less than unity \citep{huang:10,zorec:16}. While our numerical implementation is far from perfect, it is clear that centrifugal mass loss does begin at sub-critical rotation rates and likely intensifies as the breakup limit is approached.

%In Figure \ref{HR_wind} we demonstrate Hertzsprung--Russell diagrams of stars with centrifugal mass loss and stellar winds. It is clear that most mass loss results from stellar winds while centrifugal mass loss only accounts for a very small fraction. Especially for massive stars, centrifugal mass loss only accounts for less than $1\times 10^{-7}$ of the total mass loss.

\begin{figure}
\begin{center}
\includegraphics[scale=0.53]{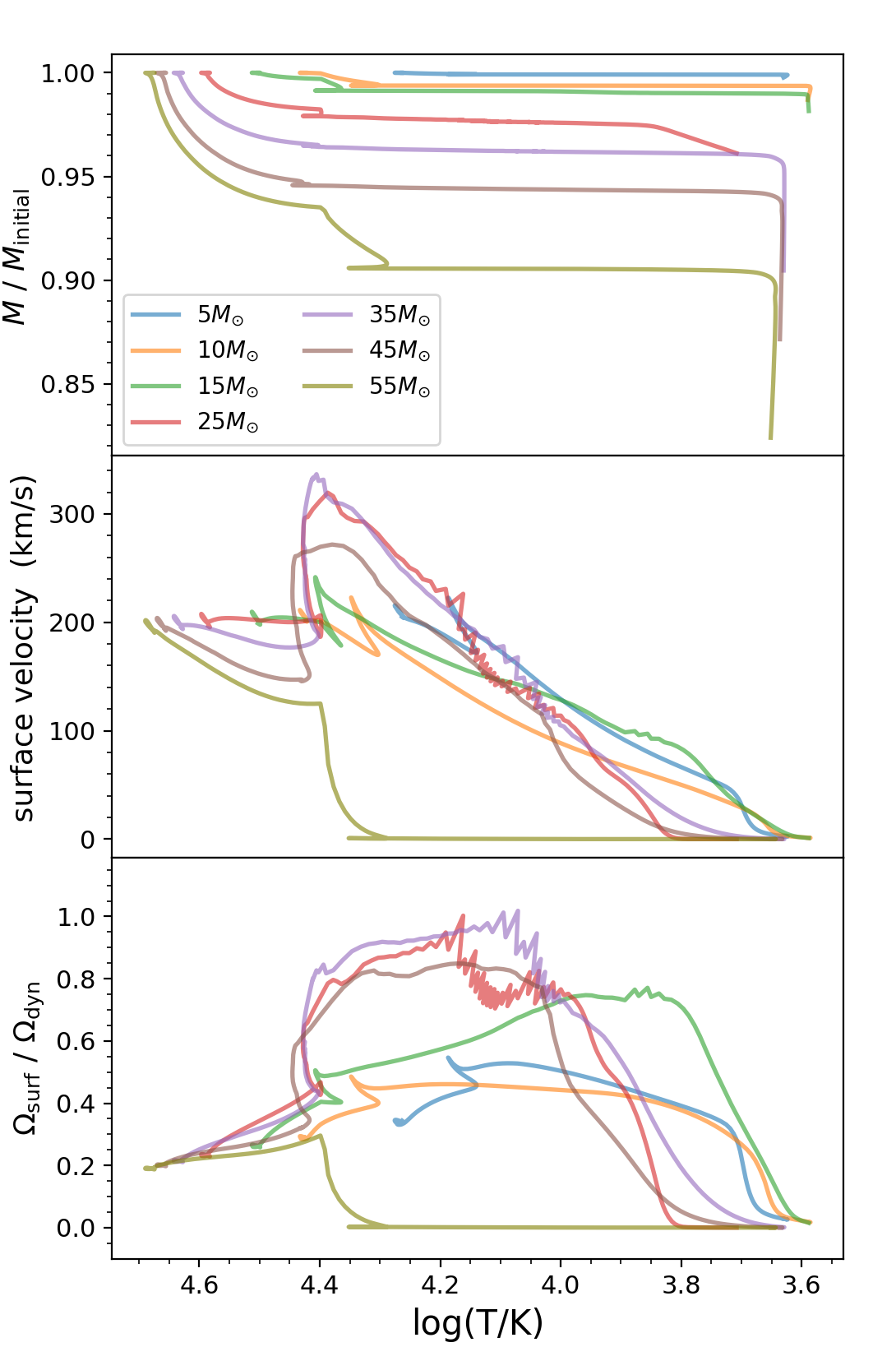}
\end{center}
\caption{\label{mass_low_metallicity}
Same as Figures \ref{mass_nowind} and \ref{mass_wind}, but for stars with half-solar metallicity ($Z = 0.0085$). Compared to solar metallicity, the mass loss rates are significantly lower, allowing the stars to retain more angular momentum, such that centrifugal mass loss is more pronounced.
}
\end{figure}

In Figure \ref{HR_wind}, we present Hertzsprung--Russell diagrams of stars with centrifugal mass loss and stellar winds, with points again colored by mass loss rate. It is clear that most mass loss results from stellar winds, while centrifugal mass loss only accounts for a very small fraction. However, even in these models, the centrifugal mass loss does become substantial for $\sim 20 \, M_\odot$ stars crossing the Hertzsprung gap. Centrifugal mass loss is most prominent in the mass range $\sim 15-25 \, M_\odot$ because lower mass stars experience less evolution-induced spin-up, and higher mass stars lose too much AM through winds. While centrifugal mass loss is negligible compared to the total mass lost, it may cause stellar outbursts in this part of the HR diagram (see Section \ref{instability}) if centrifugal mass loss occurs episodically. Interestingly, this mass loss is largest (and would likely produce more violent outbursts) in stars near the S Doradus instability region, and near the sgB[e] stars, both of which are known to exhibit outbursts that eject mass.

\subsection{Effects of Metallicity}

\begin{figure}
\includegraphics[scale=0.34]{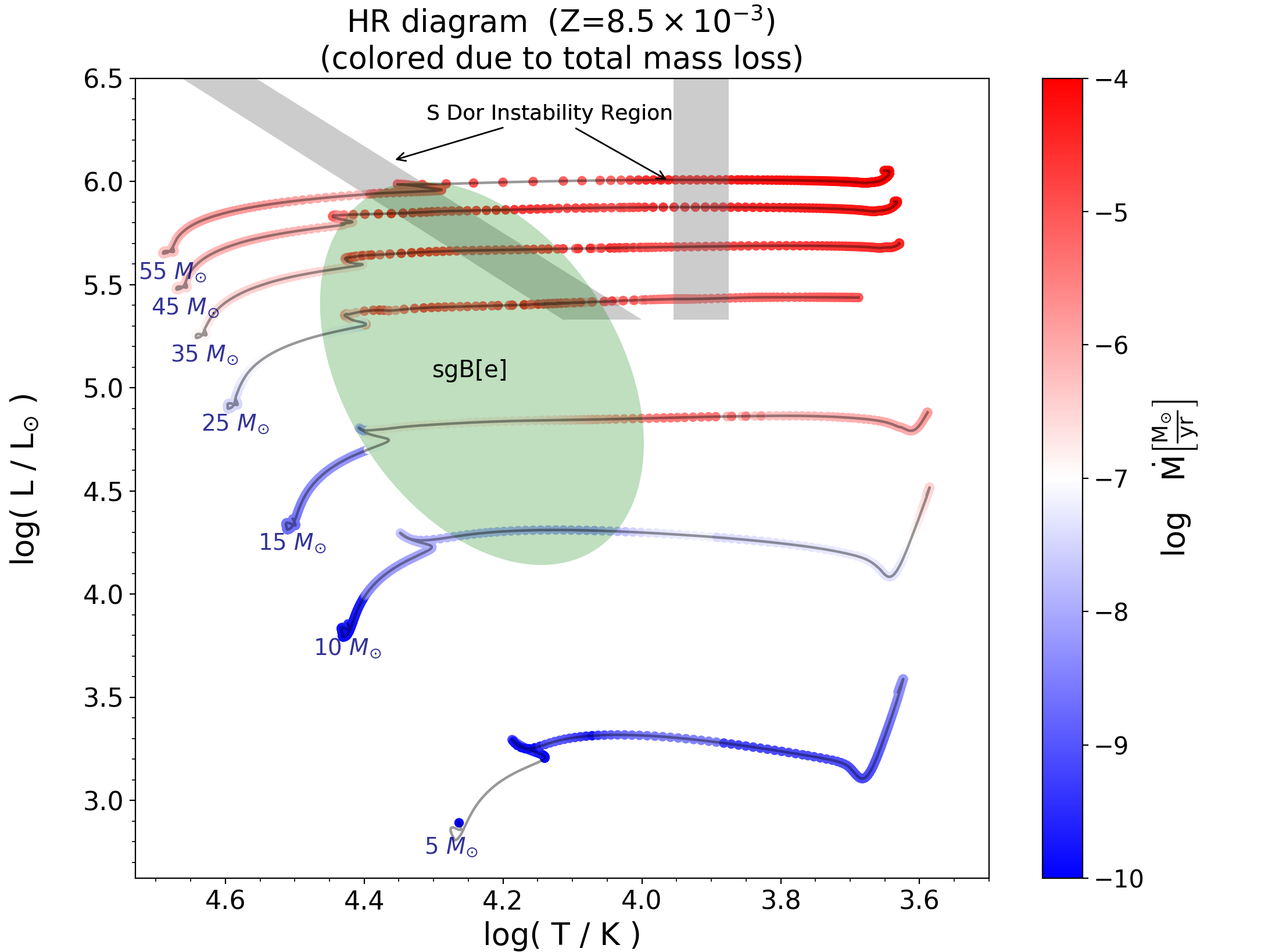}
\hspace{0cm}
\includegraphics[scale=0.34]{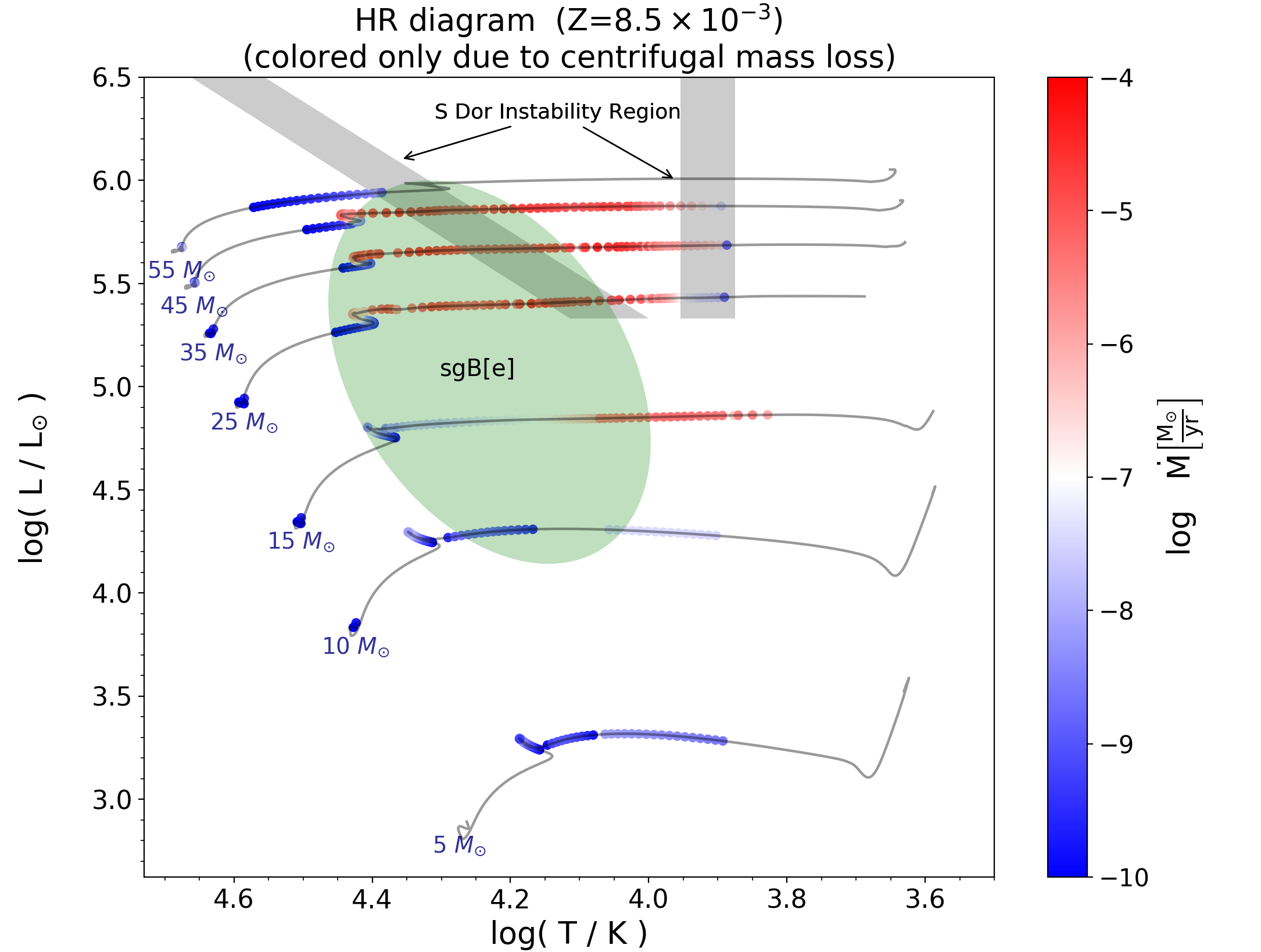}
\caption{\label{lowmetal_wind}
Same as Figure \ref{HR_wind}, but for stars with half-solar metallicity. Centrifugal mass loss is much more common for these models, frequently occurring near the sgB[e] and S Doradus instability regions.
}
\end{figure}

%To demonstrate the shifting emphasis from stellar winds to centrifugally driven mass loss in low-metallicity regions, we construct models with initial metallicity half of the previous ones, i.e. the initial metal mass fraction $Z=8.5\times 10^{-3}$.

Massive star winds are sensitive to stellar metallicity since they are driven by absorption lines of iron group elements. Hence, we expect lower metallicity stars to lose less mass and AM via line-driven winds, potentially increasing the importance of centrifugal mass loss. To demonstrate the shifting emphasis from stellar winds to centrifugally driven mass loss in low-metallicity regions, we construct models with initial metallicity half of the previous ones, i.e., $Z=8.5\times 10^{-3}$.

%Figure \ref{mass_low_metallicity} illustrates the evolution of mass and surface rotational velocity of our low-metallicity models. We can see that, compared to Figure \ref{mass_wind}, their surface rotational velocities drop down slower and can even spin up because weaker line-driven effects  allow the stars to maintain more AM. Subsequently, more preserved AM fuels centrifugal effects and drives the stars to rotate closer to the break-up limit.

Figure \ref{mass_low_metallicity} illustrates the evolution of mass and surface rotational velocity of our low-metallicity models, and Figure \ref{lowmetal_wind} shows the centrifugal mass loss rates. We can see that, compared to solar-metallicity stars, the surface rotation velocities of low-metallicity stars remain larger because winds remove much less AM. The higher AM fuels much larger centrifugal mass loss rates, especially for high-mass stars. We see high post-MS centrifugal mass loss extending up to $\approx 45 \, M_\odot$, compared to only $\approx 25 \, M_\odot$ for solar metallicity. The centrifugal mass loss also occurs over a broader span of the HR diagram, but remains largest as stars cross the HR gap, near the locations of sgB[e] and S Doradus stars. While centrifugal mass loss remains a small fraction of the total mass loss, it can dominate the mass loss rate for post-MS B/A/F-type supergiants evolving into red supergiants.

%Such an intensification in centrifugal mass loss from low-metallicity stars is obvious by contrast between Figure \ref{HR_wind} and Figure \ref{lowmetal_wind}. From these Hertzsprung--Russell diagrams, one can clearly see the prominently increasing centrifugal mass loss rate in low-metallicity stars.

\begin{figure}
\includegraphics[scale=0.56]{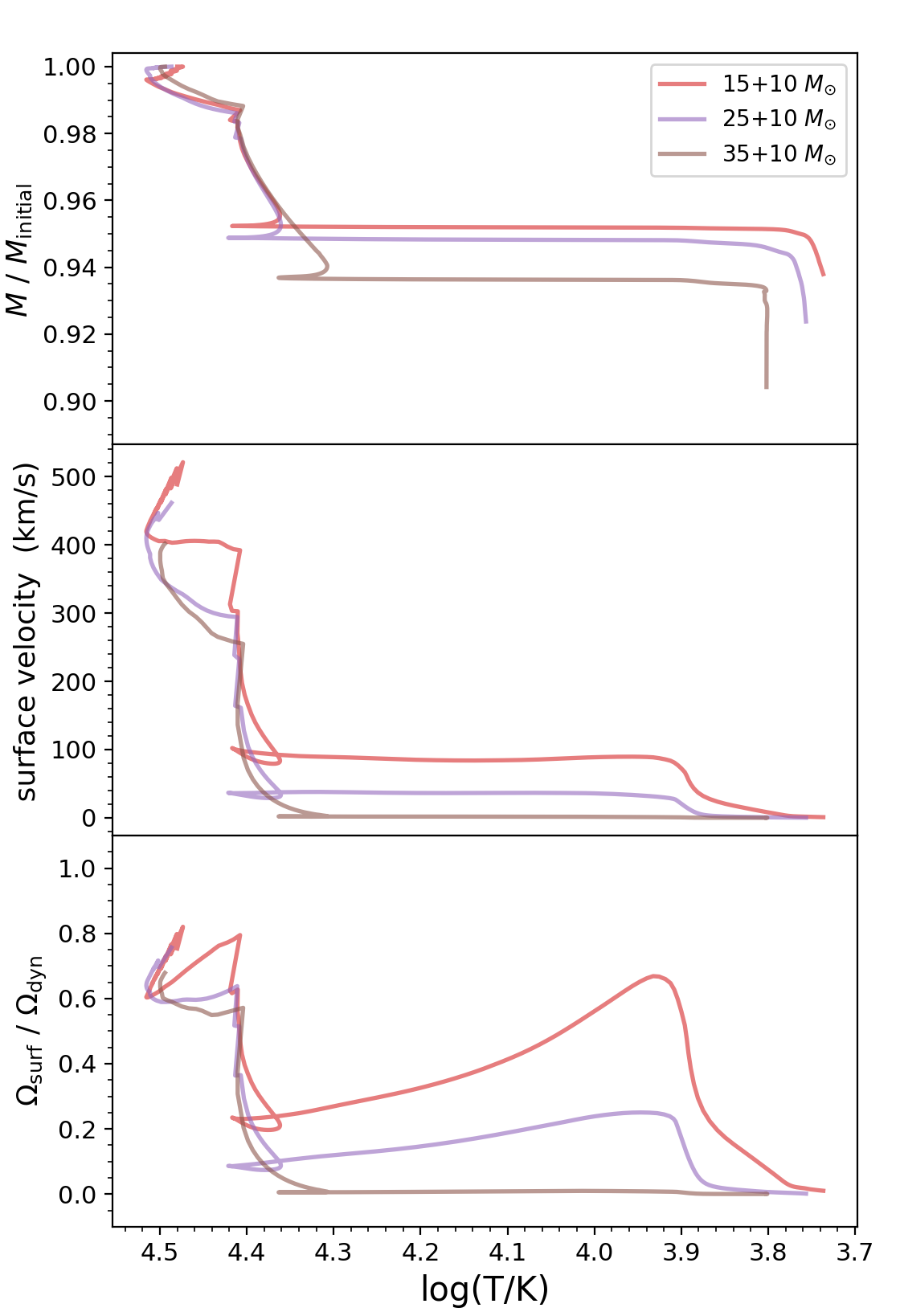}
\caption{\label{mass_binary_+10} Same as Figures \ref{mass_nowind}, \ref{mass_wind} and \ref{mass_low_metallicity}, but for stars that accrete 10$M_{\odot}$ near the end of the main sequence. Tracks begin at the end of mass accretion. Compared to single stars, the rotation rate is larger at the end of hydrogen burning, intensifying centrifugal effects. }
\end{figure}

These results suggest that centrifugal mass loss should be much more prominent in low-metallicity environments. If the formation of Be stars, sgB[e] stars, and LBV outbursts is linked to centrifugal mass loss, we expect these phenomena to be more common in low-metallicity galaxies such as the LMC and SMC. For Be stars, this appears to be the case (e.g., \citealt{peters:19}). In particular, these single-star models predict more high-mass, high-temperature, and high-luminosity Be stars in low-metallicity environments, whereas the increase for lower mass stars is more mild. While these lower mass stars dominate Be rates due to the steep initial mass function, the mass-sensitivity of centrifugal mass loss may help distinguish it from binary Be star formation models.

\subsection{Binary Interaction}
\label{binary}

\begin{figure}
\includegraphics[scale=0.32]{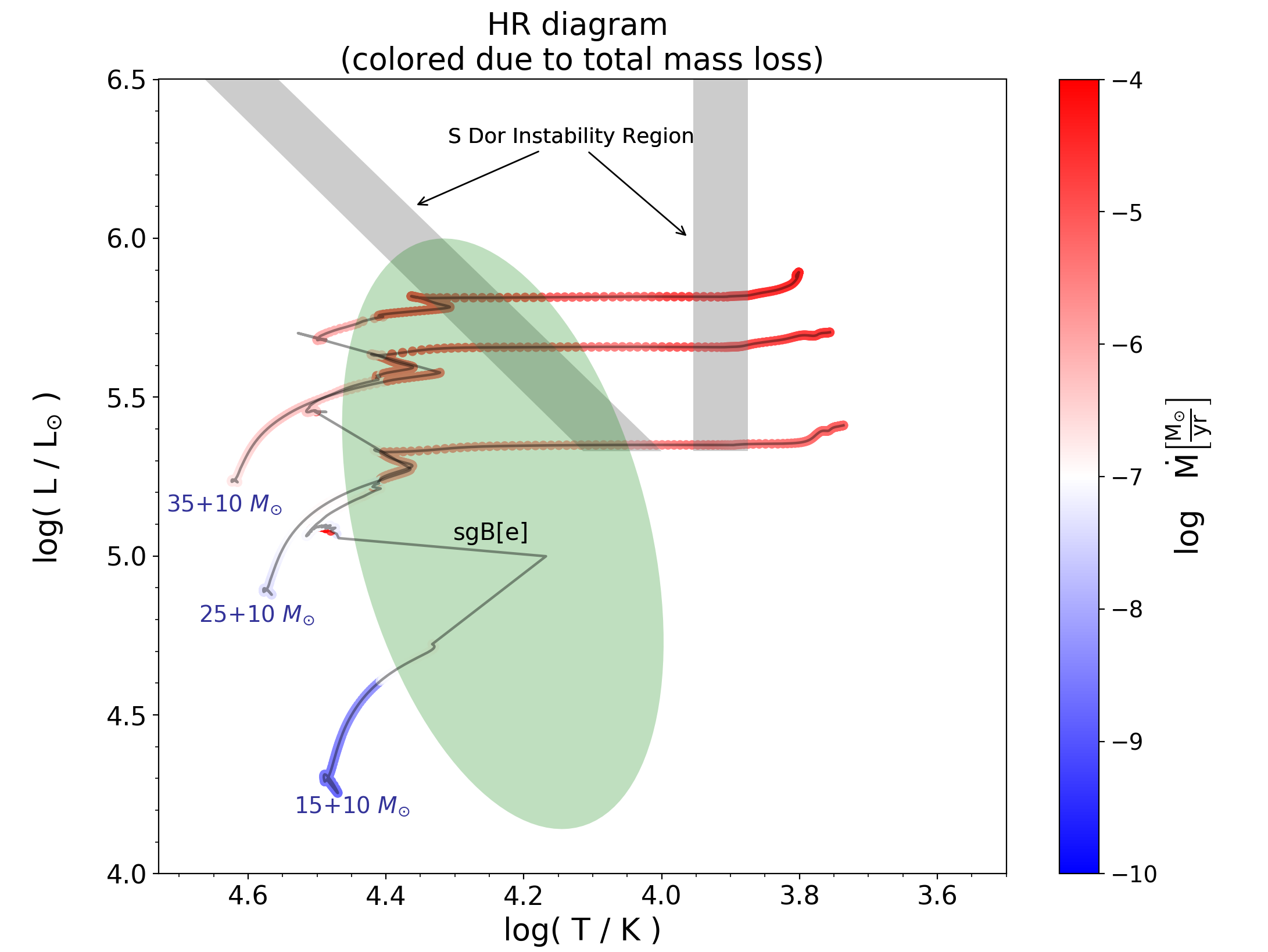}
\hspace{0cm}
\includegraphics[scale=0.32]{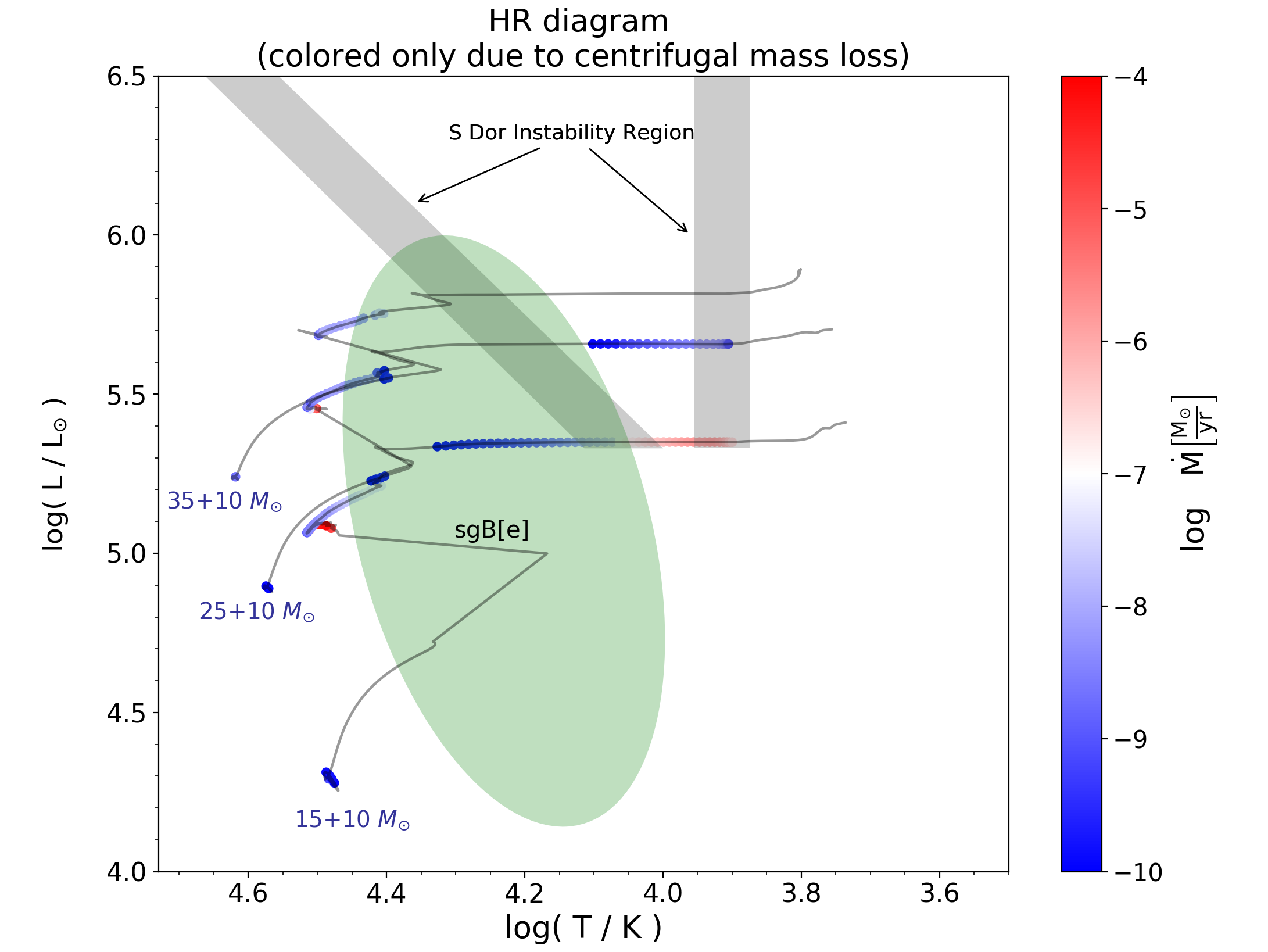}
\caption{\label{HR_binary_+10}
%Same as Figure \ref{HR_wind}, but for stars with 10$M_{\odot}$ mass accretion when they evolve from central H burning to central He burning. Centrifugal mass loss rates are larger for these models, frequently occurring near the sgB[e] and S Doradus instability regions
Same as Figure \ref{HR_wind}, but for stars that accrete 10 $M_{\odot}$ near the end of the main sequence. Centrifugal mass loss is slightly more pronounced for these models due to their increased angular momentum when they evolve off the main sequence.
}
\end{figure}

%Although for single-star models centrifugal mass loss is almost negligible compared with stellar winds, centrifugal effect has increasing significance in binary systems. When a massive star merges with its companion, the accreted mass and AM can suddenly drive the star to spin up largely due to the conservation of AM, and then trigger centrifugal mass loss. 

Centrifugally driven mass loss may be more pronounced in binary systems where the primary star has accreted mass via stable mass transfer or a stellar merger. The accretion event can replenish the AM of the star, causing it to spin much more rapidly at the end of the MS, intensifying centrifugal mass loss.

%To simulate such a binary scenario, we construct our MESA model by exerting a mass accretion via mass\_change instruction when the star evolves from core H burning to core He burning. After the star has gained $10M_{\odot}$ mass, we stop accretion and impose a rotation rate close to the break-up limit, then let the star evolve with both stellar wind and centrifugal mass loss until central helium mass fraction is below 10\%.

To simulate such a binary scenario in our MESA models, we evolve them until the central hydrogen mass fraction is $X=1\%$, and then add $10 \, M_\odot$ using \verb|relax_mass|. This mimics a stellar merger as the star expands near the end of the MS. After the mass change, we use \verb|relax_omega_div_omega_crit| to adjust the surface rotation rate until $\Omega_{\rm surf} / \Omega_{\rm crit}=1$, which is reasonable because a star need only gain $\sim \kappa M$ in order to spin up to breakup, and from Figure \ref{rotation} we see that typically $\kappa \! \sim \! 10^{-2}$ near the end of the MS. We then let the star evolve with both stellar wind and centrifugal mass loss until the central helium mass fraction is below 10\%.

Implementing the aforementioned binary prescriptions, we run models with initial masses of 15, 25, and 35 $M_{\odot}$. Figure \ref{mass_binary_+10} displays the evolution of mass and surface rotational velocity, beginning after the accretion event. In each of these models, the hydrogen-burning core grows and ingests new hydrogen after accretion, so that the star returns to the MS, behaving similar to a MS star of the new mass. However, because much of the core hydrogen has already been burnt, the models spend less time on the MS after accretion, and they lose less AM to winds by the time they evolve off the MS again. Hence, upon leaving the MS, our binary models have greater surface velocities and are closer to the break-up limit, intensifying centrifugal mass loss.

%We demonstrate the mass loss rate stemming from centrifugal effect in Figure \ref{HR_binary}, and from the HR diagram we can see centrifugal mass loss rates as large as $\sim 10^{-7}\ M_{\odot}/{\rm yr}$ and $\sim 10^{-5}\ M_{\odot}/{\rm yr}$ are achieved for the 25+10 $M_{\odot}$ and 15+10 $M_{\odot}$ models separately. Compared with Figure \ref{HR_wind}, this directly shows the significance of centrifugal mass loss in binary systems.

Figure \ref{HR_binary_+10} shows HR diagrams of our binary models. We can see that centrifugal mass loss rates of $\sim \! 10^{-5}\ M_{\odot}/{\rm yr}$ are achieved for the 15+10 $M_{\odot}$ model as it crosses the Hertzsprung gap, similar to the $25 \, M_\odot$ model in Figure \ref{HR_wind}. The more massive models lose more AM via winds, but the 25+10 $M_\odot$ model exhibits more centrifugal mass loss than a 35 $M_\odot$ single star due to the replenished AM described above. Hence, stellar merger events that occur near the end of the MS can generate more rapidly rotating stars that are more likely to exhibit centrifugally driven outbursts.

\begin{figure}
\includegraphics[scale=0.56]{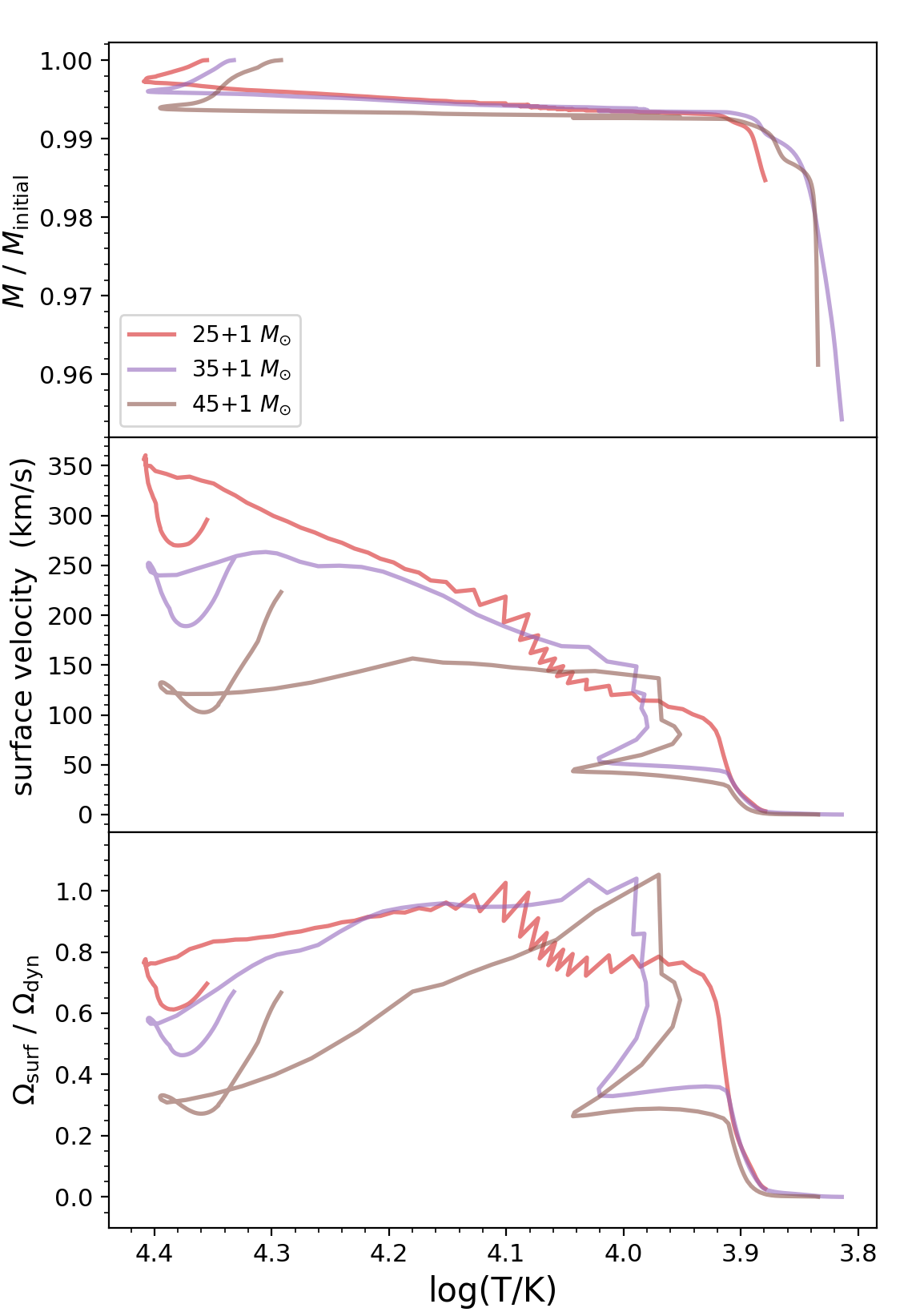}
\caption{\label{mass_binary_+1} Same as Figures \ref{mass_nowind}, \ref{mass_wind}, \ref{mass_low_metallicity} and \ref{mass_binary_+10}, but for stars that accrete 1$M_{\odot}$ near the end of the main sequence. Tracks begin at the end of mass accretion. Compared to single stars, the rotation rate is larger at the end of hydrogen burning, intensifying centrifugal effects. }
\end{figure}

Massive stars have small moments of inertia, with $\kappa \sim 10^{-2}$ near the end of the MS (Figure \ref{rotation}), such that only small amounts of mass accretion can greatly increase their spin. A large increase in the AM of the primary star $\Delta J \sim m J_{\rm crit}/(\kappa M)$ can be achieved after accreting a small mass $m$, and a $m \! \sim \! 1\,M_{\odot}$ merger suffices to spin up the primary to the break-up limit and trigger centrifugal effects. To demonstrate the centrifugal mass loss generated by ingesting a small companion, we run models with the same ingredients as described above, except that they have initial masses of 25, 35, and 45 $M_{\odot}$ and they gain 1 $M_{\odot}$ during mass transfer.

Figure \ref{mass_binary_+1} displays the evolution of these models. Compared with Figure \ref{mass_wind} and \ref{mass_binary_+10}, the post-MS ratio $\rm{\Omega_{surf}/\Omega_{dyn}}$ substantially increases, making centrifugal mass loss much more important. The reason is that ingesting only $1 \, M_\odot$ spins the primary up to breakup, but it hardly increases the MS lifetime, so the primary is unable to lose much mass through stellar winds after the accretion event. These stars thus enter the post-MS with much more AM, making them more prone to centrifugal mass loss when the core contracts and the star crosses the HR diagram.

\begin{figure}
\includegraphics[scale=0.34]{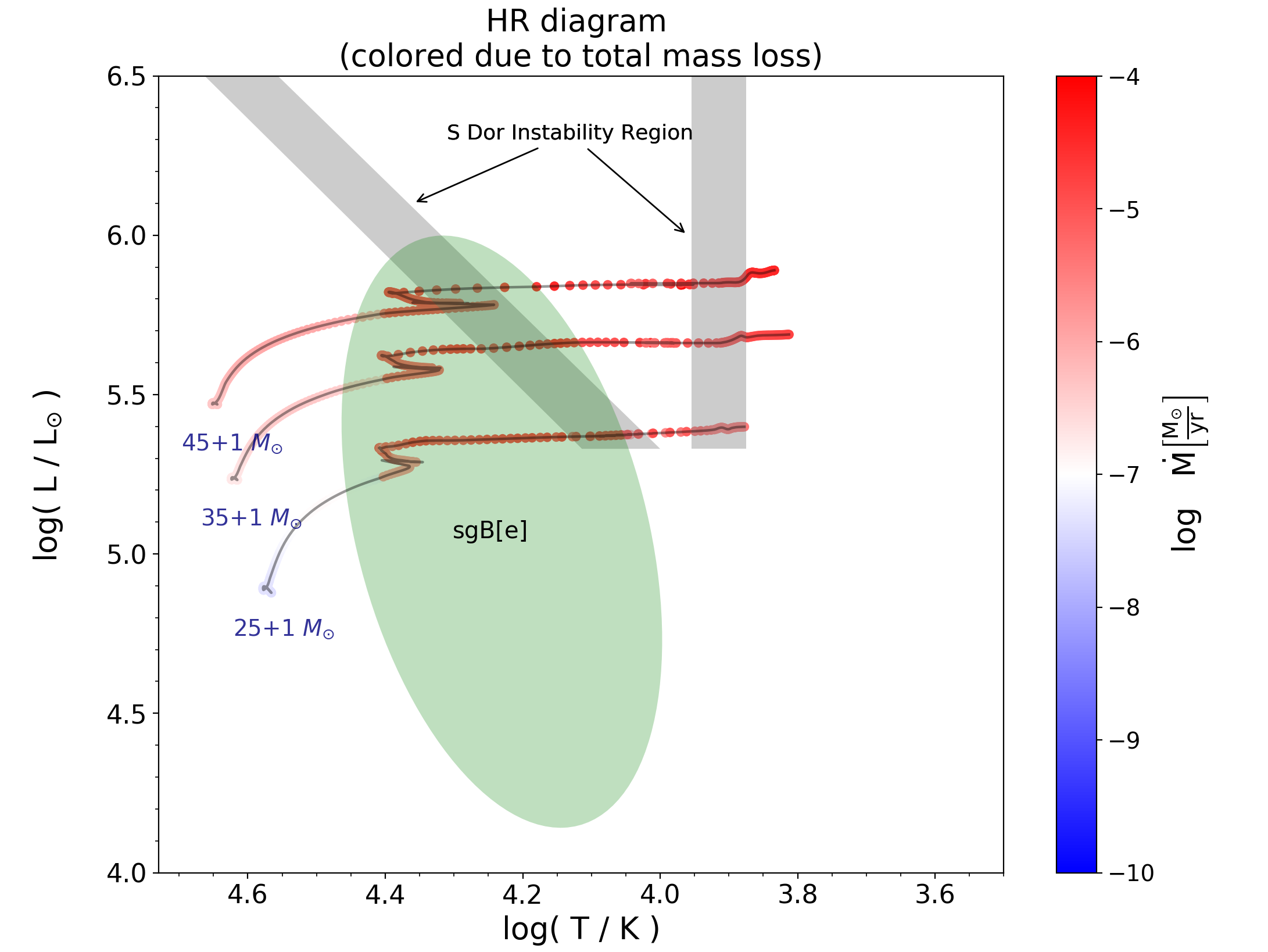}
\hspace{0cm}
\includegraphics[scale=0.34]{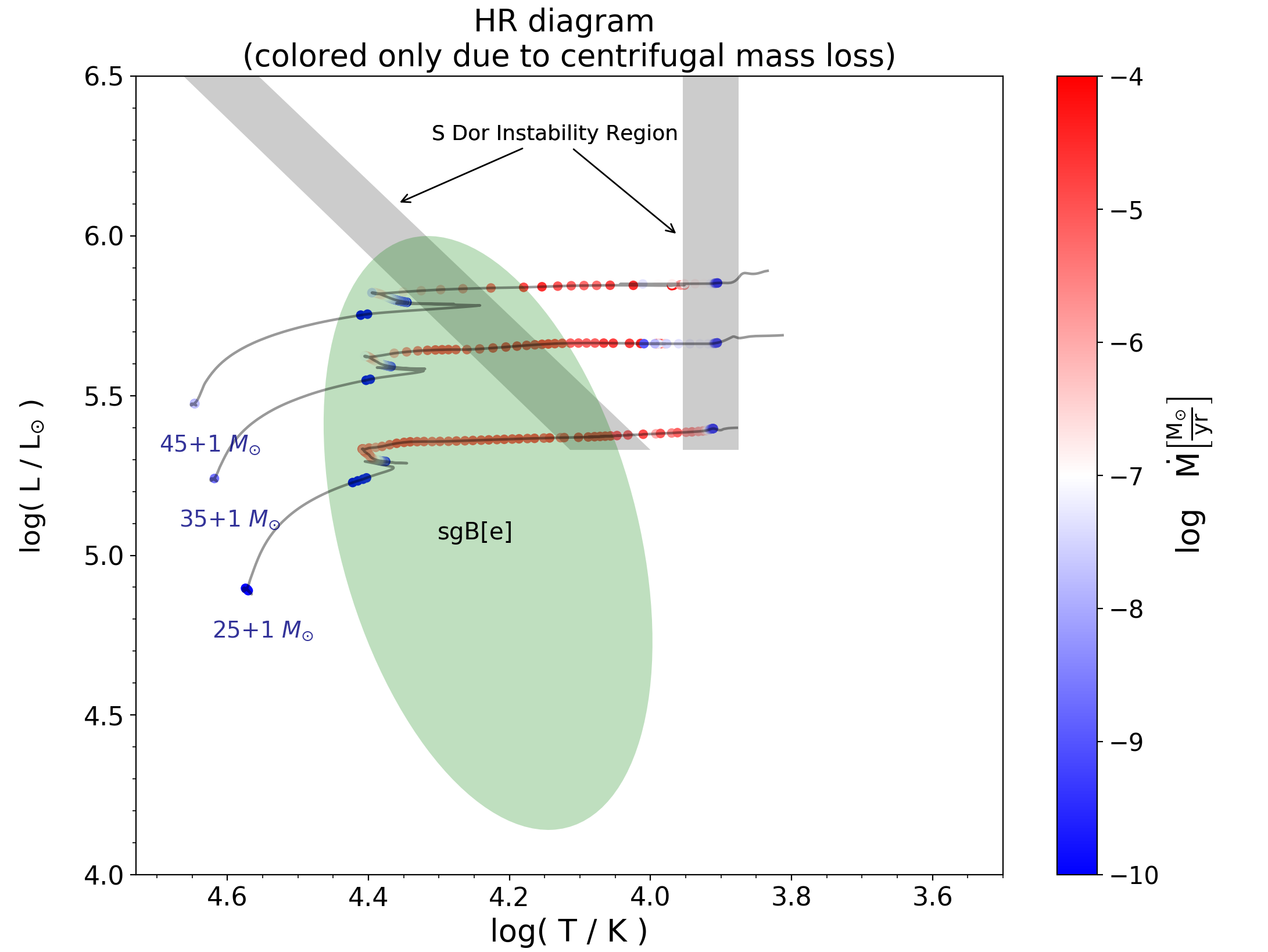}
\caption{\label{HR_binary_+1}
Same as Figure \ref{HR_wind} and \ref{HR_binary_+10}, but for stars that accrete 1 $M_{\odot}$ near the end of the main sequence. Centrifugal mass loss rates are larger for these models due to their increased angular momentum when they evolve off the main sequence.
}
\end{figure}

From the HR diagram in Figure \ref{HR_binary_+1}, it is clear that small amounts of mass accretion greatly intensify centrifugal mass loss rates of all models. The effect is most prominent for the $45\,M_{\odot}$ model, where large centrifugal mass loss rates of greater than $10^{-5}\ M_{\odot}/{\rm yr}$ are achieved within the S Doradus instability region. In contrast, centrifugal effects are nearly negligible when there is no mass gain, due to AM loss by winds (Figure \ref{HR_wind}). Hence, assimilating a small companion at the end of MS can fuel vigorous centrifugal mass loss during the post-MS, which could play a role in the variability or outbursts of some LBVs.

\section{Centrifugal Instability}
\label{instability}

%Another way to take into account the centrifugal effects is from a perturbed point of view. For the perturbation of a thin shell with radius $r$ inside the star, such that the perturbed radial coordinate is:
%\begin{align}\label{ptb}
%    r'=r+\Delta r
%\end{align}
%and satisfies:
%\begin{align}\label{ptbr}
%    \Delta r \propto e^{-i\omega t}
%\end{align}
%we derive the expression of $\omega^2$, the square of angular frequency. As soon as $\omega^2<0$, the perturbation becomes unstable since $e^{-i\omega t}$ exponentially increases over time with $\omega$ being an imaginary number and this could potentially drive a mass loss. Before we start our derivation, we first introduce a factor $f_{\Omega}$ describing the magnitude of AM transport inside a star and calculate the radial component of centrifugal force $f_{\rm cen}$.

When centrifugal mass loss occurs, it could occur steadily, or in sporadic outbursts. To investigate the latter possibility, we perform a stability analysis of perturbations in rotating stars, including the centrifugal force. In our simplified calculation, we focus on adiabatic quasi-radial perturbations. Consider the perturbation of a shell with radius $r$ inside the star, such that the perturbed radial coordinate is:
\begin{align}\label{ptb}
    r'=r+\Delta r
\end{align}
with time-dependence
\begin{align}\label{ptbr}
    \Delta r \propto e^{-i\omega t} \, .
\end{align}
Unstable perturbations occur when $\omega^2<0$, and the existence of growing modes signals an instability that could result in mass loss through outbursts.

%Perturb the specific angular momentum $j=\frac{2}{3}\Omega r^2$ and we get:
%\begin{align}\label{ptbspecificam}
%    \frac{\Delta j}{j}=2\frac{\Delta r}{r}+\frac{\Delta \Omega}{\Omega}
%\end{align}
%where $\Omega$ is a shellular rotation rate. $f_{\Omega}$ is defined as:
%\begin{align}\label{fomega}
%    \frac{\Delta \Omega}{\Omega}=f_{\Omega}\frac{\Delta r}{r}
%\end{align}
%For conserved specific angular momentum $\Delta j=0$ , we have $f_{\Omega}=-2$, while for efficient AM transport we have $\Delta \Omega=0$ hence $f_{\Omega}=0$.
%The average radial component of the centrifugal force over a shell is computed as:
%\begin{align}\label{fcen}
%    f_{\rm cen}=\frac{1}{4{\pi}r^2}\int_0^{\pi}2\pi \Omega^2 r^3 sin^3\theta d\theta=\frac{2}{3}\Omega^2r
%\end{align}

The Lagrangian perturbation to the specific angular momentum $j=\frac{2}{3}\Omega r^2$ is
\begin{align}\label{ptbspecificam}
    \frac{\Delta j}{j}=2\frac{\Delta r}{r}+\frac{\Delta \Omega}{\Omega}
\end{align}
where $\Omega$ is a shellular rotation rate. We now introduce a factor $f_{\Omega}$ describing the change in rotation rate
\begin{align}\label{fomega}
    \frac{\Delta \Omega}{\Omega}=f_{\Omega}\frac{\Delta r}{r} \, .
\end{align}
For conserved specific AM $\Delta j=0$ , we have $f_{\Omega}=-2$, while for efficient AM transport during the pulsation, we have $\Delta \Omega=0$ hence $f_{\Omega}=0$. Usually AM is thought to be conserved, but if the plasma is threaded by a magnetic field, for example, magnetic torques could enforce nearly constant rotation and an increased specific AM (at the expense of AM from other layers). The spherically averaged radial component of the centrifugal force over a shell is computed as:
\begin{align}\label{fcen}
    f_{\rm cen} &= \frac{1}{4{\pi}r^2}\int_0^{\pi}2\pi \Omega^2 r^3 sin^3\theta d\theta \nonumber \\
    &=\frac{2}{3}\Omega^2r \, .
\end{align}
The spherically averaged perturbation to the centrifugal force is then
\begin{align}\label{ptbfcen}
    \frac{\Delta f_{\rm cen}}{f_{\rm cen}}=2\frac{\Delta \Omega}{\Omega}+\frac{\Delta r}{r}=(2f_{\Omega}+1)\frac{\Delta r}{r} \, .
\end{align}

\begin{figure}
\includegraphics[scale=0.55]{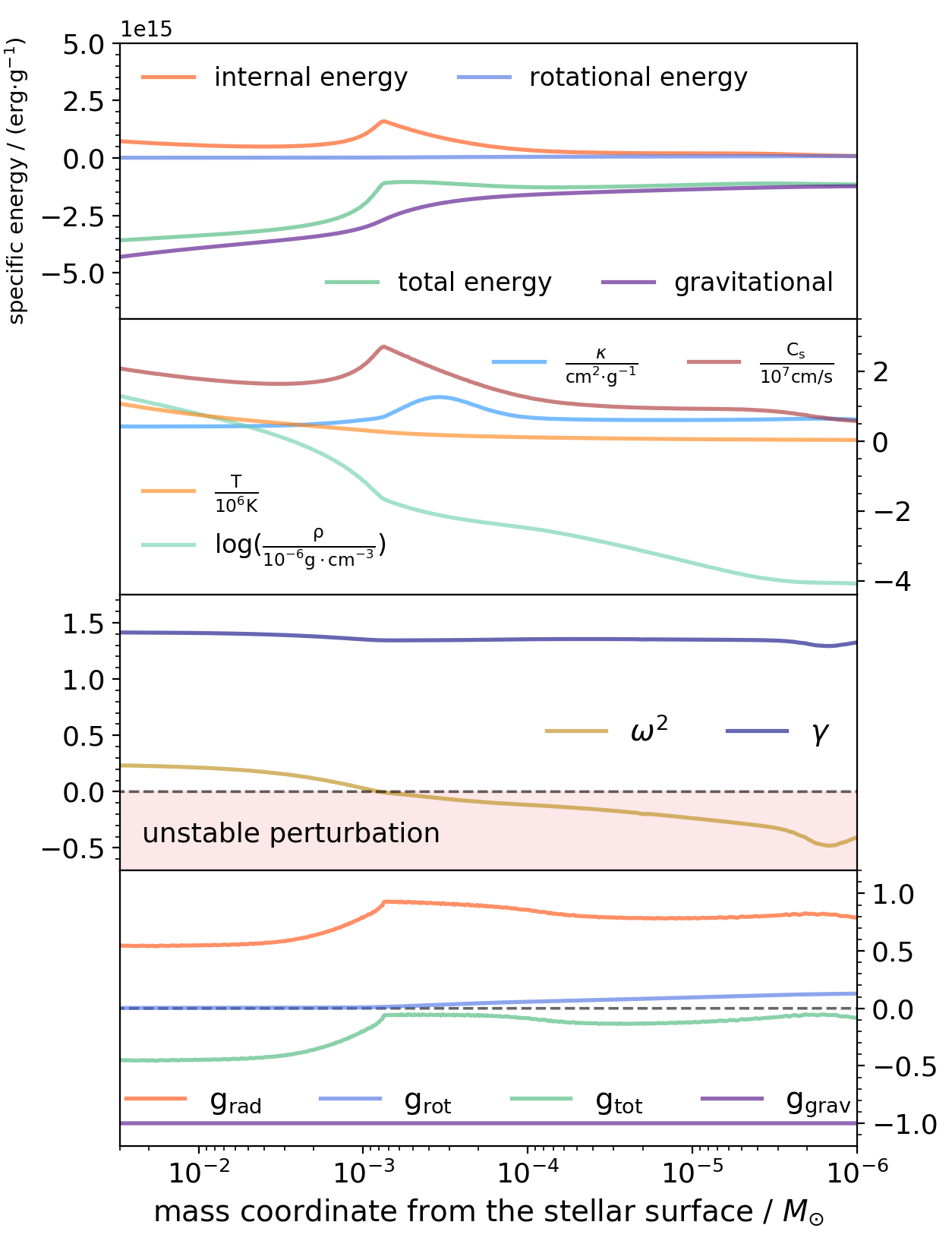}
\caption{\label{structure} Profiles of stellar properties for a $45 M_\odot$ model when its surface temperature is $T_{\rm eff} = 20000$ K. The $x$-axis is the exterior mass, i.e., the mass coordinate measured from the star's surface. {\bf{Panel 1 (top):}} Specific energies, with colors corresponding to different kinds of energy as labeled. {\bf{Panel 2:}} Opacity, sound speed, density and temperature in the outer layers of the star. Note the peaks in opacity and sound speed near the iron opacity bump at mass coordinate $\approx \!6 \times 10^{-4} \, M_\odot$. {\bf{Panel 3:}} Dimensionless square frequency $\omega^2$ of the perturbation in units of $GM/R^3$, along with the adiabatic index $\gamma$. {\bf{Panel 4 (bottom):}} Specific forces in units of $GM/R^2$.}
\end{figure}

%Now we derive the expression of $\omega^2$. As a first step we list all the relations between $\Delta r$ and other perturbed quantities. Assuming the perturbation $\Delta \vec{r}=\lambda \vec{r}$ for a small constant number $\lambda$, we have the perturbed continuity equation:
%\begin{align}\label{ptbconeq}
%    \Delta \rho +\rho\nabla\cdot\Delta\vec{r}=0
%\end{align}
%and this yields:
%\begin{align}\label{rhor}
%    \frac{\Delta \rho}{\rho}=-3\frac{\Delta r}{r}
%\end{align}
%We assume adiabatic pulsations such that:
%\begin{align}\label{pr}
%    \frac{\Delta p}{p}=\gamma\frac{\Delta \rho}{\rho}=-3\gamma\frac{\Delta r}{r}
%\end{align}
%In Figure \ref{peakxm} we make a plot to display the difference between thermal and dynamical timescales near the iron opacity peak to exemplify the reason to chose adiabatic approximation. At the iron opacity peak, the thermal timescale is generally more than $10^4$ larger than the dynamical scale, indicating that heat can not diffuse efficiently and adiabatic pulsations are properly applied.

\begin{figure}
\includegraphics[scale=0.57]{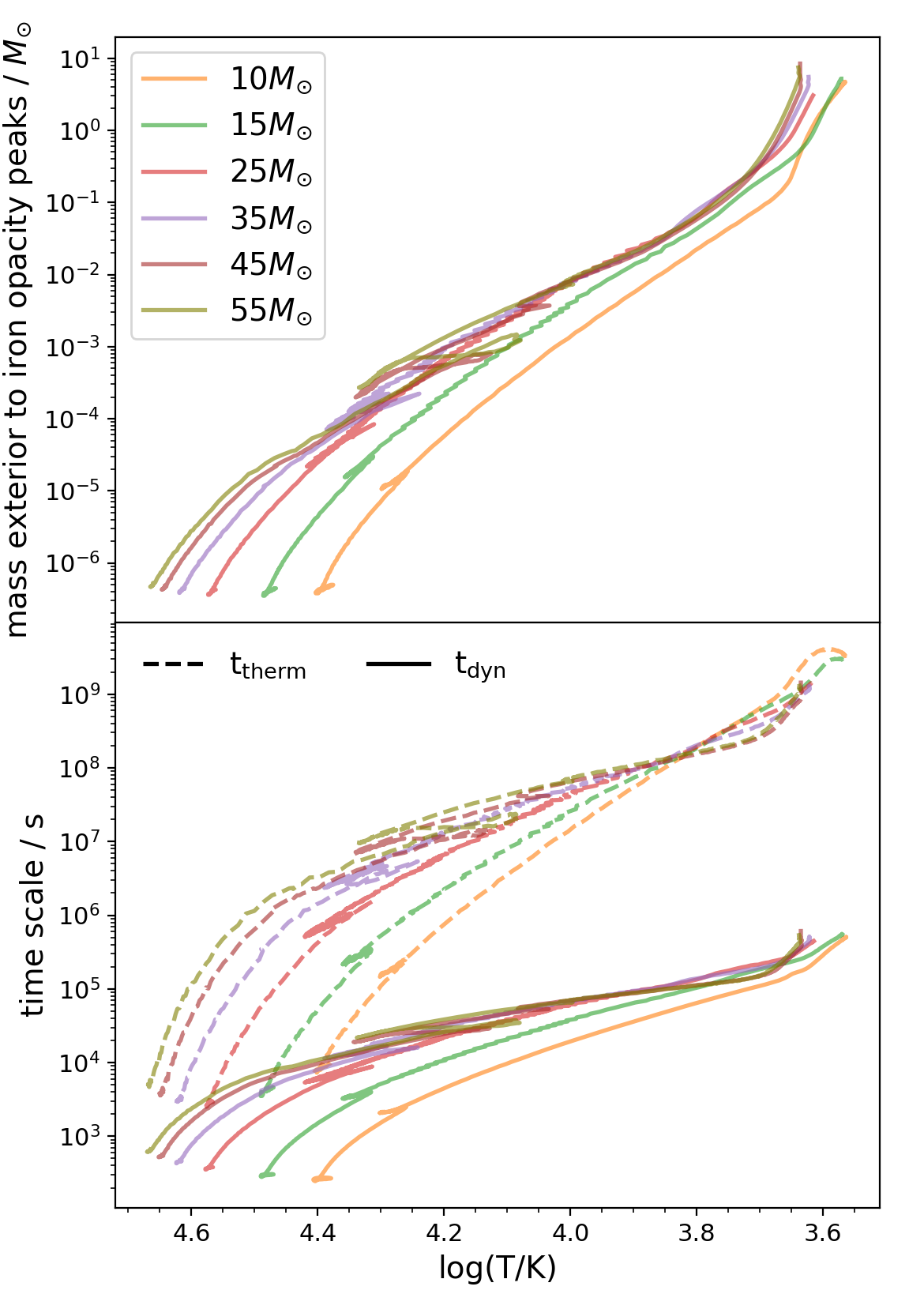}
\caption{\label{peakxm} {\bf{Top:}} The mass exterior to the iron opacity peak during the evolution of stars with initial masses 15-55 $M_{\odot}$ and initial surface rotation velocity of 200 km/s, as a function of surface temperature. {\bf{Bottom:}} Local timescales at the iron opacity peak, with dashed lines for the thermal timescale $\rm{t_{therm}}$ and solid lines for the dynamical timescale $\rm{t_{dyn}}$. It is clear that in general $\rm{t_{therm} \gg t_{dyn}}$, justifying our adiabatic approximation.}
\end{figure}

Next, we have the perturbed continuity equations
\begin{align}\label{ptbconeq}
    \Delta \rho +\rho\nabla\cdot\Delta\vec{r}=0 \, .
\end{align}
For simplicity, we consider homologous perturbations to the stars such that $\Delta r \propto r$, so that 
\begin{align}\label{rhor}
    \frac{\Delta \rho}{\rho}=-3\frac{\Delta r}{r} \, .
\end{align}
This assumption essentially restricts our analysis to modes with long wavelengths in the outer part of the star, i.e., similar to radial fundamental modes in the envelope (though potentially with more complicated dependence in the core). We also assume adiabatic pulsations such that
\begin{align}\label{pr}
    \frac{\Delta p}{p} = \gamma\frac{\Delta \rho}{\rho} = -3\gamma\frac{\Delta r}{r} \, .
\end{align}
We shall see that instability is most likely to occur near the iron opacity bump in the outer layers of the star. To justify our adiabatic approxiation, Figure \ref{peakxm} shows the thermal time scale $t_{\rm therm} = 4 \pi r^2 H \rho c_P T/L$ and dynamical timescale $t_{\rm dyn} = H/c_s$ near the iron opacity peak, where $H$ is the pressure scale height. The thermal timescale is typically a few orders of magnitude larger than the dynamical time scale, indicating that heat can not diffuse efficiently and an adiabatic approximation is reasonable. Consequently, our analysis does not include strange modes trapped in outermost layers where non-adiabatic growth/damping rates can be very large.

%The perturbed gravitational force and radial centrifugal force are separately:
%\begin{align}\label{ptbg}
%    \frac{\Delta{g_{\rm{grav}}}}{{g_{\rm{grav}}}}=-2\frac{\Delta r}{r}
%\end{align}
%\begin{align}\label{ptbfcen}
%    \frac{\Delta f_{\rm cen}}{f_{\rm cen}}=2\frac{\Delta \Omega}{\Omega}+\frac{\Delta r}{r}=(2f_{\Omega}+1)\frac{\Delta r}{r}
%\end{align}

%Here we summarize the relation between $\Delta r$ and other perturbed quantities:
%\begin{align}\label{ptbquantites}
%    \frac{\Delta r}{r}=-\frac{1}{3}\frac{\Delta \rho}{\rho}=-\frac{1}{3\gamma}\frac{\Delta p}{p}=-\frac{1}{2} \frac{\Delta{g_{\rm{grav}}}}{{g_{\rm{grav}}}}=\frac{1}{2f_{\Omega}+1}\frac{\Delta f_{\rm cen}}{f_{\rm cen}}
%\end{align}
%Next, by perturbing the radial Navier-Stokes equation, we have:
%\begin{align}\label{ptbNS}
%    \rho\frac{\partial}{\partial t}\Delta v_r=-\frac{\partial}{\partial r}\Delta p-\frac{\partial p}{\partial r}\frac{\Delta r}{r}-\Delta(\rho{g_{\rm{grav}}})+\Delta(\rho f_{\rm cen})
%\end{align}
%there, $\Delta v_r$ is the perturbed radial velocity and note that the background on which we perturb satisfies $v_r=0$.

Next, by perturbing the radial Navier-Stokes equation, we have:
\begin{align}\label{ptbNS}
    \rho\frac{\partial}{\partial t}\Delta v_r &= - \rho \omega^2 \Delta r -\frac{\partial}{\partial r}\Delta p \nonumber \\ 
    &+\frac{\partial p}{\partial r}\frac{\Delta r}{r}-\Delta(\rho{g_{\rm{grav}}})+\Delta(\rho f_{\rm cen})
\end{align}
where $\Delta v_r$ is the perturbed radial velocity, and the background on which we perturb satisfies $v_r=0$. Note we have ignored the Coriolis force, which we justify in Appendix \ref{coriolis} under certain conditions.

The perturbed gravitational force is:
\begin{align}\label{ptbg}
    \frac{\Delta{g_{\rm{grav}}}}{{g_{\rm{grav}}}}=-2\frac{\Delta r}{r} \, .
\end{align}
To compute $\frac{\partial}{\partial r}\Delta p$, recall in equation \ref{ptbquantites} we have $\frac{\Delta p}{p}\propto \frac{\Delta r}{r}= {\rm constant}$, and
\begin{align}\label{approx}
    \frac{\partial}{\partial r}\Delta p=p\frac{\partial}{\partial r}\left(\frac{\Delta p}{p}\right)+\frac{\Delta p}{p}\frac{\partial p}{\partial r} \nonumber \\
    \approx \frac{\Delta p}{p}\frac{\partial p}{\partial r}
\end{align}
The last line follows from our assumption that $\Delta p/p$ varies slowly, whereas $p$ varies over a pressure scale height that is typically much smaller than $r$ near the iron opacity peak.

%To compute $\frac{\partial}{\partial r}\Delta p$, recall in (\ref{ptbquantites}) we have $\frac{\Delta p}{p}\propto \frac{\Delta r}{r}= {\rm constant}$, and thus we make the following approximation:
%\begin{align}\label{approx}
%    \frac{\partial}{\partial r}\Delta p=p\frac{\partial}{\partial r}\left(\frac{\Delta p}{p}\right)+\frac{\Delta p}{p}\frac{\partial p}{\partial r}\approx \frac{\Delta p}{p}\frac{\partial p}{\partial r}
%\end{align}
%Plug equations (\ref{ptbr}), (\ref{ptbquantites}) and (\ref{approx}) into (\ref{ptbNS}), and calculate $\frac{\partial p}{\partial r}$ from the hydrostatic background, we finally have the expression of $\omega^2$:
%\begin{align}\label{omega2}
%    \omega^2=\frac{g_{\rm{grav}}}{r}\left[3\gamma-4-\left(2f_{\Omega}+3\gamma-1\right)\frac{f_{\rm cen}}{g_{\rm{grav}}}\right]
%\end{align}
%In Figure \ref{structure} we show a profile of a $45M_{\odot}$ star with an effective temperature $T_{\rm eff}=20,000$ K. The peak of internal energy is almost at the same location where the perturbation becomes unstable, i.e. $\omega^2<0$, which both can possibly distablize the star. Since the internal energy peak results from the opacity peak associated with iron group element, we make Figure \ref{peakxm} to show the mass exterior to this peak, which is considered unstable.

To summarize, the relation between $\Delta r$ and other perturbed quantities is
\begin{align}\label{ptbquantites}
    \frac{\Delta r}{r} &= -\frac{1}{3}\frac{\Delta \rho}{\rho}=-\frac{1}{3\gamma}\frac{\Delta p}{p} \nonumber \\ &= -\frac{1}{2} \frac{\Delta{g_{\rm{grav}}}}{{g_{\rm{grav}}}} = \frac{1}{2f_{\Omega}+1}\frac{\Delta f_{\rm cen}}{f_{\rm cen}} \, .
\end{align}
From hydrostatic equilibrium, $\frac{\partial p}{\partial r} = - \rho g_{\rm grav} + f_{\rm cen}$. Then, plugging in equations \ref{ptbr}, \ref{approx}, and \ref{ptbquantites} into \ref{ptbNS}, we finally obtain an expression for $\omega^2$:
\begin{align}\label{omega2}
    \omega^2=\frac{g_{\rm{grav}}}{r}\left[3\gamma-4-\left(2f_{\Omega}+3\gamma-1\right)\frac{f_{\rm cen}}{g_{\rm{grav}}}\right] \, .
\end{align}

Equation \ref{omega2} is similar to the classic result for non-rotating stars, which become less stable as $\gamma$ decreases toward $4/3$, but with an extra term proportional to the centrifugal force. We note that our local result is very similar to the global result derived in \cite{ledoux:45} (see also \citealt{tassoul:78}), in the limit that AM is conserved ($f_\Omega = -2$), in which case the centrifugal term is actually stabilizing for $\gamma < 5/3$ as we expect in massive stars. However, our formula allows for the destabilizing influence of torques that conserve rotation rate instead of AM ($f_\Omega =0$), in which case the centrifugal term is destabilizing. In this case, instability occurs even at relatively small values of $f_{\rm cen}/g_{\rm grav}$, as small as $f_{\rm cen}/g_{\rm grav} = 1/4$ for $\gamma = 5/3$, and even smaller values for $\gamma < 5/3$.
%In Figure \ref{structure} we show a profile of a $45M_{\odot}$ star with an effective temperature $T_{\rm eff}=20,000$ K. The peak of internal energy is almost at the same location where the perturbation becomes unstable, i.e. $\omega^2<0$, which both can possibly distablize the star. Since the internal energy peak results from the opacity peak associated with iron group element, we make Figure \ref{peakxm} to show the mass exterior to this peak, which is considered unstable.

In Figure \ref{structure}, we examine the structure of a $45M_{\odot}$ star with an effective temperature $T_{\rm eff}=20,000$ K. The iron opacity peak occurs at a mass coordinate of $q \approx 4 \times 10^{-4} \, M_\odot$, and is also associated with a peak in internal energy density, sound speed, and outward radiation force, as well as a dip in $\gamma$. At mass coordinates above this location, we see the local value of $\omega^2$ becomes negative, indicating an instability could originate from this location in the star. While the value of $\omega^2$ is even more negative closer to the surface of the star, our adiabatic approximation could break down there.

Our perturbation analysis is crude and not suitable for robust predictions. The main point is to highlight the destabilizing influence of the centrifugal force for any perturbation in which AM can be transported rapidly (i.e., shorter than an oscillation time scale). Since the fundamental mode frequencies of $\gamma = 4/3$ polytropes approach zero in the absence of rotation, their long pulsation periods will allow for more AM transport during the pulsation cycle, further destabiliizing them relative to polytropes of higher $\gamma$. The growth of such a perturbation is fueled by energy added to the outwardly perturbed fluid element via these (presumably magnetic) torques. Of course, the star must conserve its total AM, so the inner layers must contract (or at least expand by a smaller fraction) in order for the outer layers to gain energy and AM during the pulsation cycle. Hence, the pulsation mode cannot everywhere have the $\Delta r \propto r$ dependence we have assumed, but must have more complex spatial structure, as we should expect for realistic stars with kinks in their density profiles (Figure \ref{peakxm}).

Crucially, this centrifugal instability does not require super-Eddington luminosities or near-breakup rotation at the surface of the star, which are the criteria usually examined for massive stars. Instead, the instability is promoted by a low adiabatic index and high rotation rate inside the star, and it feeds on the internal energy and rotational energy of sub-surface layers. The instability is most likely to occur in the outer layers of post-MS massive stars where $\gamma$ approaches $4/3$ over a significant fraction of the stellar radius, unlike some MS models where the iron opacity bump is located at a radius very close to the photosphere. Hence, instabilities causing outbursts are more likely to occur in very luminous stars post-MS such as LBVs, and are less likely to occur in MS stars such as Be stars. 

We have performed exploratory one-dimensional hydrodynamic simulations with MESA, including centrifugal forces and AM transport. These simulations do indeed give rise to an instability near the end of the MS, approximately when we find $\omega^2 < 0$ near the iron opacity bump. The instability only occurs in rotating models, but it occurs even at moderate rotation rates well below breakup. Hence we believe the instability is essentially the same as that predicted by equation \ref{omega2}. The outcome can be a steady centrifugally driven wind, or a periodic expansion and contraction of the stellar envelope. While such expansion/contraction does occur in S Doradus outbursts, it typically occurs on a timescale of $\sim$weeks in these models, much faster than the slow observed variability of S Doradus cycles. The instability may saturate when the expanding envelope expands and cools, increasing its adiabatic index and the value of $\omega^2$. A detailed investigation requires much more work, but these preliminary numerical calculations (which are global and include non-linear effects) do indicate the presence of a centrifugally driven instability. 

If an instability does occur, the outcome is not clear, but it could result in centrifugal ejection of material overlying the iron opacity peak. Figure \ref{peakxm} plots the mass exterior to the iron opacity peak for stars evolving across the HR diagram. Where centrifugally driven mass loss is most likely to occur, i.e., $\log T \approx 4-4.3$, the associated mass is $\sim 10^{-5}-10^{-2} \, M_\odot$. While centrifugal instability may be able to power Be star outbursts and S Doradus outbursts, it likely cannot power giant eruptions from LBVs that eject more than $\approx \! 1 \, M_\odot$ of material.

\section{Discussion}
\label{discussion}

The centrifugal effect included in our models is an important factor in stellar evolution since many massive stars will reach the break-up limit when they evolve off the MS. We find that such an effect is greater for more massive and lower metallicity stars, especially when they are crossing the Hertzsprung gap. At this stage, centrifugal mass loss rates as large as $10^{-4}M_{\odot}/{\rm yr}$ are achieved as shown in Figures \ref{HR_nowind}, \ref{HR_wind}, and \ref{lowmetal_wind}. However, for massive stars in high-metallicity regions with strong line-driven winds, centrifugal mass loss becomes less important because stellar winds carry away most of the star's AM, preventing the star from reaching the break-up limit during the post-MS. Below we discuss implications and caveats of our findings.

%Our centrifugal mass loss mechanism has important implications for LBVs and Be stars. In Figure \ref{HR_nowind}, we see that centrifugal mass loss rates are usually largest near the S Doradus instability strip, where LBVs exhibit sporadic outbursts and eruptions. This indicates that LBVs can undergo violent centrifugal mass loss which may potentially trigger massive eruptions at this stage. It is also notable that prominent centrifugal effects take place near the super-giant B[e] stars, implying efficient centrifugal mass ejections around the equators of Be stars which may cause the circumstellar disks. We investigate centrifugal instabilities from a perturbed point of view. 

\subsection{The Be Star Phenomenon}

Be stars are known to be very rapidly rotating, often near their breakup velocities \citep{rivinius:13}, but their formation is not totally understood. Our work suggests that some Be stars can be produced as a result of core contraction and surface rotational acceleration due to efficient AM transport. This single star scenario, proposed by \cite{maeder:01}, also explains the higher abundance of Be stars in slightly aged populations (older than 10 Myr) and stars near the TAMS \citep{keller:00,aidelman:18}, because more envelope spin-up occurs as stars age and evolve toward the TAMS. Additionally, the larger abundance of Be stars in low-metallicity environments (\citealt{peters:19} and references therein) is naturally explained by the smaller mass/AM loss through winds (rather than having to appeal to larger rotation rates at birth), allowing more stars to reach their breakup rotation rate. 

\cite{ekstrom:08} predicted the number of stars that reach breakup via this channel during the MS and then compare with the observed fraction of Be stars in clusters at various metallicities. However, they underpredict the observed fraction of Be stars by a factor of a few. While a few effects could account for this discrepancy, an important difference is that their AM transport prescription only accounts for Eddington-Sweet circulation and hydrodynamic turbulence. Consequently, their models (and those in most papers by the Geneva group) exhibit significant differential rotation on the MS, with surface rotation reduced by inefficient AM transport from core to surface. \cite{ekstrom:08} agrees that inclusion of magnetic torques via the Tayler-Spruit (TS) dynamo \citep{spruit:02} would increase AM transport and Be star formation. In the late stages of preparation of this manuscript, \cite{hastings:19} examined Be star formation in models including AM transport via the TS dynamo. They indeed find a higher rate of Be star formation, approximately consistent with observations if Be star are formed when $v_{\rm rot}/v_{\rm crit} \gtrsim 0.75$. 
%The mass loss prescriptions included in \cite{hastings:19} are different than ours,

Since we now know that AM transport in radiative regions of stars is even more efficient than the TS dynamo predicts \citep{cantiello:14}, a larger fraction of Be stars should be produced. Because our AM transport prescription and the TS dynamo predict nearly rigid rotation during the MS, we expect to find similar results to \cite{hastings:19} for MS stars, though we have not performed a full population synthesis calculation. Like \cite{hastings:19}, we find a 'goldilocks' range of $\approx 15-25 \, M_\odot$ (at solar metallicity) where Be star formation is most efficient, the lower bound arising from a smaller core mass and the upper bound from larger wind AM losses. The upper bound extends to larger mass/luminosity in low-metallicity environments. If AM is transported in radiative zones via fossil fields \citep{kissin:15}, AM transport is even more efficient than our estimates, further promoting Be star formation. Our results show that Be formation is most likely to occur in stars just beyond the MS and could also contribute to the population of B[e] supergiants \citep{kraus:19}. However, because stars evolve across the HR gap extremely quickly, post-MS Be stars will likely be outnumbered by MS Be stars.

Some aspects of Be stars remain difficult to explain with our single star models. For instance, like \cite{hastings:19}, our single star models predict that Be stars are much more common near the TAMS, whereas only a mild increase is observed. Additionally, we may expect higher levels of nitrogen enrichment than observed. Furthermore, we cannot readily explain the higher incidence of Be stars in massive clusters \citep{bastian:17} and the apparent bimodal spin distribution, which may require binary and/or environmental effects. All of this evidence may point to a large fraction of Be stars arising from binary channels in which the Be progenitor accreted mass/AM from a companion star. Indeed, binary models of Be star formation \citep{pols:91,shao:14} can account for a large fraction of the Be star population. \cite{demink:13} showed that the fraction of rapidly rotating post-mass transfer systems expected from binary population synthesis is similar to the observed number of Be stars. Our binary models in Section \ref{binary} show that Be star formation is certainly promoted by mass accretion, either via stable mass transfer or stellar mergers. Hence, we agree that a substantial fraction of Be stars arise from binary interactions, but we reaffirm that they can produced by single stars, especially in low-metallicity environments.

\subsection{The LBV Phenomenon}

An implication of our work is that centrifugally driven mass loss could be related to LBV outbursts, especially S Doradus variations. Indeed, the centrifugal mass loss rates we predict are similar to LBV mass loss rates of a few $10^{-5} \, M_\odot$/yr during S Doradus outbursts \citep{smithrev:14,smith:17}. Intriguingly, we predict the highest mass loss rates for stars near the S Doradus instability region (Figure \ref{HR_nowind}), because this is when the cores of very massive stars are contracting rapidly and providing AM to the surface layers, driving rapid centrifugal mass loss. While measuring the rotation rates of LBVs is difficult, \cite{groh:06,groh:09} has indeed demonstrated that some LBVs exhibit very rapid rotation, near their breakup rates, consistent with our suggestion of centrifugally driven outbursts.

We note that our models predict post-MS centrifugally driven mass loss to be most common for solar metallicity at masses of $\sim 15-25 \, M_\odot$, or luminosities of $\sim \! 10^{5} \, L_{\odot}$, near or below the bottom of the LBV region shown in Figure \ref{HR_wind}. In light of the lower masses and luminosities recently measured for some LBVs \citep{smith:15,smithlbv:19,smith:19}, the instability region should be revised downward. Since our models do not rely on super-Eddington mass loss, these low-luminosity LBVs could naturally be explained by the centrifugal mass loss processes we have investigated. \cite{smith:19} also showed that some LBVs lie to the left of the classical S Doradus instability strip, a feature also seen in many of our models (e.g., Figures \ref{lowmetal_wind} and \ref{HR_binary_+1}). Our models also predict high-luminosity LBVs to be more common in low-metallicity environments where wind mass loss is less important.

In this picture, LBVs may simply represent stars at the luminous end of a continuum of stars exhibiting centrifugal mass loss. This continuum could extend from Be stars at luminosities of just $\sim \! 10^2 \, L_\odot$, through sgB[e] stars with luminoisties of $\sim \! 10^5 \, L_\odot$, all the way to LBVs with $L \gtrsim 10^6 \, L_\odot$. The observational characteristics may be quite different, especially since the massive end of this continuum has near-Eddington luminosities, while radiative acceleration is negligible for ordinary Be stars. Hence, we may not expect to see centrifugally expelled decretion disks around LBVs like we do in Be stars, because such disks would quickly blown away by radiation force, or they would collapse due to 'Poynting-Robertson' drag \citep{owocki:98} on lines of iron-group elements.

The centrifgual mass loss and/or outburst process may also be quite different in stars of different types. In MS Be stars, the iron opacity peak is relatively near the photosphere, where the overlying mass is small ($\lesssim 10^{-5} \, M_\odot$, Figure \ref{peakxm}) and the thermal time is small, promoting shorter and less violent (or non-existent) outbursts. In contrast, for post-MS LBV stars, the mass of the iron opacity peak can be much deeper, where the overlying mass is larger ($\sim \! 10^{-2} \, M_\odot$) and the thermal time is large, promoting longer and more violent outbursts. Outbursts of more massive stars may also be more common and more violent due to the role of radiation pressure, though a detailed prediction of centrifugally driven outbursts in LBVs awaits future work. For now, it remains difficult to explain the giant eruptions of LBVs like $\eta$ Carina, which can expel more than $10 \, M_\odot$ of material \citep{smith:03,smith:07}.  While the iron opacity peak can extend to such depths in yellow/red supergiants, this occurs when the surface convection zone becomes very deep, the star's moment of inertia and spin period increases rapidly, and centrifugally driven mass loss is not expected to occur.

\subsection{Case B Stellar Mergers}

In Section \ref{binary}, we showed how case A mass accretion near the end of the MS (as could result from a stellar merger) can intensify centrifugal mass loss when the star evolves off the MS. It would also be interesting to investigate case B mass accretion that can occur when the star crosses the Hertzsprung gap and can swallow a small companion. \cite{justham:14} performs a detailed investigation of the evolution of such merger products (not including updated AM transport prescriptions). Unlike our case A merger products that resemble single stars of increased mass that usually evolve into red supergiants, case B mass accretion produces stars that typically spend the rest of their lives as blue/yellow supergiants residing near the LBV instability strip. These stars do not evolve into red supergiants with large moments of inertia, and their cores continue to contract as they evolve. Hence, they could potentially exhibit centrifugal mass loss throughout helium-burning until core-collapse. Centrifugal mass loss from such stars could produce some LBV stars, and it could contribute to pre-SN outbursts of type IIn progenitor stars, some of which are known to resemble LBVs (e.g., \citealt{gal-yam:09,smith:10,mauerhan:13,ofek:13b,ofek:14,smith:14,smithrev:14,elias-rosa:16,tartaglia:16}).

\subsection{Comparison with Observed Rotational Velocities}
\label{obs}

It is useful to compare our predictions with observed rotation rates of evolving massive stars, and with prior stellar models. Several studies \citep{huang:06a,hunter:08,dufton:13} have indicated a range of initial rotation rates for MS massive stars, with typical rotation rates in the range $0-250$ km/s, and a small tail extending to higher velocities. Our models with 200 km/s are thus fairly typical massive stars, or perhaps slightly more rapidly rotating than average.

Measurements of post-MS rotation rates are more sparse but very informative, and have been made both in the LMC and SMC \citep{dufton:06a,mcevoy:15} and in the Galaxy \citep{fukuda:82,verdugo:99,abt:02,abt:03,huang:06b}. These works find fairly small rotational velocities for most A/B-type supergiant stars (which presumably have just evolved off the MS), with many in the range 0-100 km/s, and decreasing rotation towards lower temperature/gravity. For relatively low-mass ($5-15 M\odot$) sub-giants, nearly solid body rotation and nearly constant rotational velocity on the MS appears to match the data fairly well \citep{abt:02,huang:10}. Agreement with our models and others (e.g., \citealt{meynet:00,heger:00,ekstrom:08}) is fairly good on the MS where they predict smaller amounts of differential rotation. However, our modeled rotational evolution (Figures \ref{mass_wind} and \ref{mass_low_metallicity}) predict slightly faster rotation than typically observed after the TAMS. In post-MS stars, the surface rotational velocity falls faster than expected from solid-body rotation \citep{verdugo:99,abt:03}, and is below the predictions of our models. While our models do develop differential rotation after the TAMS, the level of differential rotation is less than most other models. 

Possible explanations for these discrepancies are that our AM transport prescription predicts too much core-envelope coupling, or that our wind prescription predicts too little mass loss. However, the works presented above point out that models with weaker AM transport and stronger winds (e.g., \citealt{meynet:00,meynet:03}) also predict rotation rates somewhat larger than observed in many cases. So, the source of the discrepancy is unclear.
%One possible explanation is that some low-gravity supergiants inferred to be post-MS stars may in fact be pre-MS stars that are still contracting and spinning up.
%Since these stars have a larger moment of inertia, their surface rotation rates could be slower than post-MS stars, especially in lower mass models where AM mass loss through winds is less important.
Some blue supergiants could be helium-burning stars on a blue loop that could have spun down via mass loss during a red supergiant phase (though blue loops can also cause fast rotation if the star did not lose much AM as a supergiant, \citealt{heger:98}). It has also been pointed out \citep{townsend:04} that in stars with significant centrifugal distortion, the spectrum is dominated by polar regions (due to gravity darkening near the equator) where the rotational velocity is small, so observed rotational velocities do not represent surface-averaged rotation speeds. Lastly, many of the studies discussed above do not include Be stars, which are the fastest rotating stars, and so those studies are biased towards the slowly rotating end of the distribution.

A final important possibility to consider is that convective regions do not rotate rigidly as assumed in nearly all stellar evolution codes, but instead exhibit strong differential rotation as suggested by \cite{kissin:15}. Massive post-MS stars begin to develop deep convective envelopes due to helium/iron opacity peaks for surface temperatures smaller than $\sim$20,000 K, or even higher temperatures for stars near the Eddington limit. If these convective zones rotate differentially, the surface rotation rate can be slower than the rotation rate at the top of the radiative region. In this case, modeled surface rotation rates may better agree with data and may remain smaller than the surface breakup rate. The internal rotation rate, however, could still exceed the breakup rate, and centrifugal mass loss could occur via instabilities as discussed in Section \ref{instability}. This could help explain why post-MS mass loss sometimes occurs via large eruptions emanating from deeper layers of the star, in contrast to the small outbursts typical of Be stars apparently originating near the surface. We hope to explore this possibility in future work.

\subsection{Model Uncertanties}

A significant uncertainty in our analysis is the structure of the outer layers of very luminous models, where convection becomes inefficient and the flux is locally super-Eddington due to iron and helium opacity bumps. Our models use MESA's MLT++ prescription for handling these layers \citep{paxton:13}, resulting in a more compact envelope than models without a similar prescription (e.g., \citealt{sanyal:17}). Neither set of models is perfect, and radiation hydrodynamics simulations \citep{jiang:15,jiang:17,jiang:18,jiang:18b} may be needed to guide the construction of better models. In any case, we suspect centrifugal mass loss will be \textit{more important} if stars develop inflated envelopes. Since these envelopes have very low masses but large radial extents (especially at high metallicity), they decrease the star's rotational breakup rate without significantly increasing the star's moment of inertia. Hence the maximum AM content $J_c$ of such stars may be smaller than our models, making them more susceptible to centrifugal mass loss.

A second uncertainty is that of line-driven wind mass loss rates. For hot stars, we adopt the widely used Vink prescription \citep{vink:01}, but reduced by a factor of two. If the appropriate reduction factor is even larger, as advocated in many recent works (see review in \citealt{smithrev:14}), then stars will lose less AM on the MS and centrifugal mass loss will become more important. The main effect of less efficient winds is similar to that of lower metallicity: an upward shift in the maximum stellar mass/luminosity to which centrifgual mass loss extends.

\section{Conclusion}
\label{conclusion}

We have examined the rotational evolution of massive stars, examining whether their outer layers can spin fast enough to centrifugally expel matter. Unlike most prior analyses (e.g., \citealt{ekstrom:08,hastings:19}), we incorporate very efficient angular momentum (AM) transport within our stellar models as required by asteroseismic observations (e.g., \citealt{mosser:12,cantiello:14,gehan:18,fuller:19}). Additionally, we extend our models past the main sequence to investigate post-main sequence centrifugal mass loss. We find that efficient AM transport from the contracting core to the expanding envelope can sometimes spin up the surface layers to breakup, necessitating centrifugal mass loss. Centrifugal mass loss is most prolific near the end of the main sequence and the early post-main sequence as massive stars cross the Hertzsprung gap. During these phases of evolution, the expanding envelope contains very little mass and the core contracts rapidly, donating enough AM to spin up surface layers towards breakup velocities.

At solar metallicity, this effect is strongest for $\sim \! 20 \, M_\odot$ stars (Figure \ref{HR_wind}), with some dependence on wind mass loss rates. Lower mass stars have smaller core masses/radii such that the core's contraction is not sufficient to spin up outer layers. Higher mass stars lose most of their AM via line-driven winds, preventing centrifugal mass loss. At low metallicity, the line-driven mass loss weakens, promoting centrifugal mass loss, especially in more massive stars (Figure \ref{lowmetal_wind}). While centrifugal mass loss may generate circumstellar 'decretion' disks and could drive outbursts, it cannot expel the entire hydrogen envelope, and radiatively driven winds or binary interactions are still responsible for the bulk of mass loss. We confirm that some Be stars can form from rapidly rotating single stars due to outward AM transport within the star. However, in many stars, accretion from a binary companion is likely crucial for supplying enough AM to drive centrifugal mass loss. Hence, both effectively single and binary stellar evolution may contribute substantially to the Be star phenomenon. Interestingly, stars that accrete a small amount of mass ($\sim 1 \, M_\odot$) near the end of the MS are most prone to post-MS centrifugal mass loss (Figure \ref{HR_binary_+1}), because the merger spins them up without significant extending the MS lifetime and hence the AM loss due to line-driven winds.

We predict the largest centrifugal mass loss rates ($\sim \! 10^{-4} \, M_\odot$/yr) for massive post-main sequence stars near the S Doradus instability strip. Hence, we posit that S Doradus variations and outbursts from some luminous blue variable stars (LBVs) are driven by centrifugal mass loss, in agreement with the apparent rapid rotation of some LBVs \citep{groh:06,groh:09}. Our prediction of extreme centrifugal mass loss extending down to masses of $\sim 20 \, M_\odot$ is similar to the lower end of LBVs observed in the Milky Way \citep{smith:19}. However, the absence of such low-luminosity LBVs in the Magellanic clouds suggest metallicity may also be an important factor in driving LBV outbursts, perhaps due to a radiation-centrifugal instability (Section \ref{instability}). Preliminary analytical and numerical calculations indicate a centrifugal instability could operate in massive stars crossing the Hertzspung gap, driven by AM transport into outwardly perturbed fluid elements. The instability can originate deep in the star (at the iron or helium opacity bumps) and could potentially drive outbursts that partially expel overlying layers. We hope to examine this instability and its relation to LBV outbursts in future work.

\section{Acknowledgments}

We thank Nathan Smith and Matteo Cantiello for very helpful suggestions. This research
is funded in part by an Innovator Grant from The Rose Hills Foundation, and the Sloan Foundation through grant FG-2018-10515.

%%%%%%%%%%%%%%%%%%%%%%%%%%%%%%%%%%%%%%%%%%%%%%%%%%

%%%%%%%%%%%%%%%%%%%% REFERENCES %%%%%%%%%%%%%%%%%%

% The best way to enter references is to use BibTeX:

\bibliographystyle{mnras}
\bibliography{LBVs} % if your bibtex file is called example.bib

% Alternatively you could enter them by hand, like this:
% This method is tedious and prone to error if you have lots of references
%\begin{thebibliography}{99}
%\bibitem[\protect\citeauthoryear{Author}{2012}]{Author2012}
%Author A.~N., 2013, Journal of Improbable Astronomy, 1, 1
%\bibitem[\protect\citeauthoryear{Others}{2013}]{Others2013}
%Others S., 2012, Journal of Interesting Stuff, 17, 198
%\end{thebibliography}

%%%%%%%%%%%%%%%%%%%%%%%%%%%%%%%%%%%%%%%%%%%%%%%%%%

%%%%%%%%%%%%%%%%% APPENDICES %%%%%%%%%%%%%%%%%%%%%

\appendix

\section{Neglect of Coriolis Force}
\label{coriolis}

We have ignored the Coriolis force in equation \ref{ptbNS}, which may not always be appropriate. However, we note that \cite{ledoux:45} eliminated the Coriolis forces by considering a frame that changes rotation rate throughout the pulsation cycle. In principle, such a frame could be used in our analysis as well.

Inclusion of the Coriolis force would add a term $-2 i \Omega \omega \rho \xi_\phi \sin \theta$ to the right-hand side, where $\xi_\phi$ is the displacement in the $\phi$-direction. The Coriolis force can thus be neglected if $\omega^2 \Delta r \gtrsim \omega \Omega \xi_\phi$.
%This approximation is valid in the limit that the frequency $\omega$ is very small (such that fluid elements move very slowly), or when $\xi_\phi$ is very small. Indeed, we will be interested in modes with $\omega^2 \approx 0$ on the verge of instability. 
One could imagine strong torques (due to magnetic tension forces or convective Reynold's stresses) that overwhelm the Coriolis force and prevent fluid elements from moving horizontally and satisfying this approximation. To see this, consider the effect of a radial magnetic field. The $\phi$ component of the momentum equation (assuming axisymmetry), is
\begin{equation}
\label{phimom}
- \omega^2 \xi_\phi \sim 2 i \omega \Omega (\xi_\theta \cos \theta - \Delta r \sin \theta) \, + \mathcal{O} (\omega_{\rm A}^2 \xi_\phi) \, .
\end{equation}
The last term in equation \ref{phimom} represents magnetic tension forces, assuming the magnetic fields and horizontal displacements vary on a length scale of $r$. For convective torques, the Alfv\'en frequency $\omega_A = B_r/\sqrt{4 \pi \rho r^2}$ should be replaced by a convective turnover frequency $\omega_{\rm con}v_{\rm con}/r$. Hence, Coriolis forces can be ignored in this equation if $\omega_{\rm A}^2 \xi_\phi \gtrsim \omega \Omega \Delta r$. Combining this requirement with that from the radial momentum equation, Coriolis forces are negligible if $\omega_{\rm A} \gtrsim \Omega$. In other words, torques due to magnetic fields or convection should be stronger than Coriolis torques to justify a spherical treatment of the problem.

Since the instability originates in the outer layers of stars, it is certainly possible that $\omega_{\rm A}$ or $\omega_{\rm con}$ will be large enough to overwhelm Coriolis forces near the surface of the star. In these layers, $\omega_{\rm A}$ and $\omega_{\rm con}$ become large as the density becomes small, whereas $\Omega$ is nearly constant. In the Sun, for example, both $\omega_{\rm A}$ and $\omega_{\rm con}$ are a couple orders of magnitude larger than $\Omega$ near the surface. While both $\omega_{\rm A}$ and $\omega_{\rm con}$ decrease deeper in the star, the shallow density gradient near the surfaces of massive stars (Figure \ref{peakxm}) helps them remain large. Additionally, for stars near the Eddington limit or near $\gamma=4/3$, the rotation rate $\Omega$ can be much less than breakup when the instability sets in, decreasing the Coriolis force. Third, due to the very high luminosities and low densities of the envelopes, the convective velocities approach the sound speed deep in the star (e.g., near the iron opacity bump) increasing the magnetic/convective torques. 

\section{MESA MODEL INLISTS}
\label{mesa}

We construct our stellar models using the MESA stellar evolution code \cite{paxton:11,paxton:13,paxton:15,paxton:18,paxton:19} version 11701. The inlist is as follows:
\subsection{\rm MESA Inlist}
\begin{verbatim}
&star_job
            
      pgstar_flag = .true.

      relax_initial_Z = .true.
      new_Z = 0.017d0

      new_rotation_flag = .true.
      change_rotation_flag = .true.
      set_initial_surface_rotation_v = .true.
      new_surface_rotation_v = 200

/ ! end of star_job namelist

&controls
   
      use_other_wind = .true.
      fitted_fp_ft_i_rot = .true.
      w_div_wcrit_max = 0.7

    !--------------------------------  Convergence 

      okay_to_reduce_gradT_excess = .true.
      gradT_excess_age_fraction = 0.999d0
      gradT_excess_max_change = 0.01d0 

      timestep_factor_for_retries = 0.8
      timestep_factor_for_backups = 0.8
      min_timestep_factor = 0.9
      max_timestep_factor = 1.2d0
      backup_hold = 10
      retry_hold = 3
      redo_limit = -1
      relax_hard_limits_after_retry = .false.

      newton_iterations_limit = 7
      max_model_number = 30000
      max_number_retries = 5000

    ! Fixing the position of the Lagrangian region
    ! of the mesh helps
    ! convergence near the Eddington limit
      max_logT_for_k_below_const_q = 100
      max_q_for_k_below_const_q = 0.995 
      min_q_for_k_below_const_q = 0.995 
      max_logT_for_k_const_mass = 100
      max_q_for_k_const_mass = 0.99
      min_q_for_k_const_mass = 0.99

    !extra spatial resolution
      max_dq = 0.02

    fix_eps_grav_transition_to_grid = .true.

    ! extra controls for timestep
    ! these are for changes in mdot at the onset of 
    ! mass transfer
      delta_lg_star_mass_limit = 1d-3 
      delta_lg_star_mass_hard_limit = 2d-3 
    ! these are to properly resolve core hydrogen
    ! depletion
      delta_lg_XH_cntr_limit = 0.04d0
      delta_lg_XH_cntr_max = 0.0d0
      delta_lg_XH_cntr_min = -4.0d0
      delta_lg_XH_cntr_hard_limit = 0.06d0 

    ! this is mainly to resolve properly when the star 
    ! goes off the main sequence
      delta_HR_limit = 0.01d0 
    ! relax default dHe/He, otherwise growing He core 
    ! can cause things to go at a snail pace
      dHe_div_He_limit = 2.0
    ! we're not looking for much precision at the 
    ! very late stages
      dX_nuc_drop_limit = 5d-2

      !--------------------------------  Rotation

      am_nu_ST_factor = 0
      smooth_nu_ST = 5
      smooth_D_ST = 5
      use_other_am_mixing = .true.

      am_time_average = .true.
      premix_omega = .true.
      recalc_mixing_info_each_substep = .true.
      am_nu_factor = 1
      am_nu_non_rotation_factor = 1d0
      am_nu_visc_factor = 1
      am_nu_ES_factor = 1
      angsml = 0.0

      am_D_mix_factor = 3.33d-2
      D_ES_factor = 1

      ! this is to avoid odd behaviour when a star 
      ! switches from accreting to mass losing
      max_mdot_jump_for_rotation = 1d99

      !------------------------------------  MAIN
       
      initial_mass = 35
      initial_z = 0.02
      use_Type2_opacities = .true.
      Zbase = 0.017d0

      predictive_mix(1) = .true.
      predictive_superad_thresh(1) = 0.005
      predictive_avoid_reversal(1) = 'he4'
      predictive_zone_type(1) = 'any'
      predictive_zone_loc(1) = 'core'
      predictive_bdy_loc(1) = 'top'

      dX_div_X_limit_min_X = 1d-4
      dX_div_X_limit = 5d-2
      dX_nuc_drop_min_X_limit = 1d-4
      dX_nuc_drop_limit = 5d-2
      
      !------------------------------------  WIND

      hot_wind_scheme = 'Dutch'
      cool_wind_RGB_scheme = 'Dutch'
      cool_wind_AGB_scheme = 'Dutch'
      RGB_to_AGB_wind_switch = 1d-4
      Dutch_scaling_factor = 0.0 !0.5
      mdot_omega_power = 0.0d0

      !----------------------------  OVERSHOOTING
     
      overshoot_f_above_nonburn_core = 0.025
      overshoot_f0_above_nonburn_core = 0.005
      overshoot_f_above_nonburn_shell = 0.025
      overshoot_f0_above_nonburn_shell = 0.005
      overshoot_f_below_nonburn_shell = 0.025
      overshoot_f0_below_nonburn_shell = 0.005

      overshoot_f_above_burn_h_core = 0.025
      overshoot_f0_above_burn_h_core = 0.005
      overshoot_f_above_burn_h_shell = 0.025
      overshoot_f0_above_burn_h_shell = 0.005
      overshoot_f_below_burn_h_shell = 0.025
      overshoot_f0_below_burn_h_shell = 0.005

      set_min_D_mix = .true.
      min_D_mix = 1d2
     
      !-------------------------------------  MISC

      photo_interval = 25
      profile_interval = 25
      max_num_profile_models = 3000
      history_interval = 1
      terminal_interval = 10
      write_header_frequency = 10
      max_number_backups = 500
      relax_max_number_retries = 2000
      max_number_retries = 4000

      !-------------------------------------  MESH     
      mesh_delta_coeff = 1.0
      varcontrol_target = 5d-4
 
/ ! end of controls namelist

\end{verbatim}

\subsection{\rm Implementation of Mass Loss Wind}
\label{numerics}

Our code in {\bf run\_star\_extras.f} implementing mass loss wind is as follows:
\begin{verbatim}

    subroutine my_other_wind(id, Lsurf, Msurf, 
    Rsurf, Tsurf, w, ierr)
        use star_def, only: star_info
        use star_lib, only: star_ptr
        integer, intent(in) :: id
        real(dp), intent(in) :: Lsurf, Msurf
        real(dp), intent(in) :: Rsurf, Tsurf
        !surface values (cgs)
        
        real(dp), intent(out) :: w     
        !wind in units of Msun/year (value is >= 0)
        integer, intent(out) :: ierr
        
        integer :: y                   
        !used for tanh smoothing
        
        real(dp) :: I, r, r_old        
        !moment of inertia
        !radius or the current model
        !radius of the previous model
        
        real(dp) :: cen_wind,omega_crit 
        !centrifugal mass loss rate
        !critical rotational velocity
        
        real(dp) :: Dutch_wind       
        real(dp) :: w1,w2,alfa,temp
        !used to calculate Dutch_wind
        
        type (star_info), pointer :: s
        call star_ptr(id, s, ierr)
        if (ierr /= 0) then ! OOPS
            return
        end if
             
        w = 0
        ierr = 0
        temp = 0
        cen_wind = 0
        Dutch_wind = 0

        I     = 
        dot_product(s%i_rot(:s%nz),s%dm(:s%nz))
        r     = exp(s% xh(s% i_lnR,1))                  
        r_old = exp(s% xh_older(s% i_lnR,1))            
        y = 5   !used for tanh smoothing

        cen_wind = &
        -(I-s% xtra1 - 3.0*I*(r-r_old)/r/2.0)
        /(r**2-I/2.0/s% mstar)
        /s% time_step/1.9892e33

        if (s%Lrad_div_Ledd_avg_surf.ge.0.639) then    
            omega_crit=
            sqrt(2.0/3.0*6.67428e-8*s%mstar/r**3)
            *(1-s%Lrad_div_Ledd_avg_surf)/0.361
        else
            omega_crit=
            sqrt(2.0/3.0*6.67428e-8*s%mstar/r**3)
        end if 
        !omega_crit depending on Eddington factor

        cen_wind = 
        0.5*(1+tanh(y*log(s%omega(1)/omega_crit)))
        *cen_wind
        !tanh smoothing

        s% xtra1 = I
        !save current moment of inertia

        if (cen_wind .lt. 0.0) then
            cen_wind = 0.0 
        end if
        !guarantee the star is losing mass
        
        if (s% Dutch_scaling_factor == 0) then     
            Dutch_wind = 0
        else if (Tsurf <= 10000.0) then
            call lowT_Dutch(Lsurf,Msurf,
            Rsurf,Tsurf,Dutch_wind,
            s%Dutch_wind_lowT_scheme)
        else if (Tsurf >= 11000.0) then
            call highT_Dutch(id,Lsurf,Msurf,
            Tsurf,Dutch_wind,ierr)
        else ! transition
            call lowT_Dutch(Lsurf,Msurf,
            Rsurf,Tsurf,w1,
            s%Dutch_wind_lowT_scheme)
            call highT_Dutch(id,Lsurf,Msurf,
            Tsurf,w2,ierr)
            alfa = (Tsurf - 10000.0)/1000.0
            Dutch_wind = (1-alfa)*w1 + alfa*w2
        end if
        !calculate Dutch_wind as MESA source code
        
        Dutch_wind = 
        s%Dutch_scaling_factor*Dutch_wind

        w = cen_wind + Dutch_wind   
        !total = centrifugal + line-driven

    end subroutine my_other_wind
\end{verbatim}

To run the \verb|my_other_wind| routine, one must employ subroutines from MESA source code. They are \verb|eval_highT_Dutch|, \verb| eval_lowT_Dutch|, and other subroutines related to them. These subroutines are in the \verb|private/winds.f90| file. Additionally, we include the TSF subroutine from \cite{fuller:19} with minor modifications to implement angular momentum transport. We use a tanh function to smooth the transition of centrifugal mass loss rate from 0 to a positive number when $\Omega>\Omega_{\rm crit}$ as:
\begin{align}
    \dot{M}_{\rm cen} = \frac{1}{2}\left(1+tanh\left[log{\left(\frac{\Omega}{\Omega_{\rm crit}}\right)}^y\right]\right)*\dot{M}
\end{align}
where $\dot{M}$ is calculated as in equation (\ref{mdot}), and in our code $y=5$. This tanh smoothing represses numerical instabilities, and we find results with different values of $y$ are similar.

%%%%%%%%%%%%%%%%%%%%%%%%%%%%%%%%%%%%%%%%%%%%%%%%%%

% Don't change these lines
%\bsp	% typesetting comment
\label{lastpage}
\end{document}